\documentclass[aps,twocolumn,pra,superscriptaddress,letterpaper,nofootinbib,longbibliography]{revtex4-2}
\usepackage{amssymb}
\usepackage{physics}
\usepackage{mathtools}
\usepackage{upgreek}
\usepackage{booktabs}
\usepackage{url}
\usepackage{float}
\AtBeginDocument{
\heavyrulewidth=.08em
\lightrulewidth=.05em
\cmidrulewidth=.03em
\belowrulesep=.65ex
\belowbottomsep=0pt
\aboverulesep=.4ex
\abovetopsep=0pt
\cmidrulesep=\doublerulesep
\cmidrulekern=.5em
\defaultaddspace=.5em
}

\usepackage{txfonts}
\usepackage[pdftex]{graphicx}
\usepackage[usenames, dvipsnames, svgnames, table]{xcolor}
\usepackage[colorlinks=true, citecolor=Blue,linkcolor=BrickRed, urlcolor=magenta]{hyperref}
\usepackage{bm}
\usepackage{placeins} 
\usepackage{setspace}
\usepackage[final]{changes}

\usepackage{siunitx}

\newcommand{\OSI}[2]{$\mathcal{O}\left(\SI{#1}{#2}\right)$}
\newcommand{\prtHz}{\per\sqrt{\mrm\Hz}}

\usepackage{cleveref}
\crefname{section}{Section}{Sections}
\crefname{figure}{Figure}{Figures}
\crefname{table}{Table}{Tables}
\crefname{equation}{Eq.}{Eqs.}

\newcommand{\qt}[1]{q_{#1}} 

\newcommand{\qdt}[1]{\dot{q}_{#1}} 

\newcommand{\A}{\mathbf{A}}
\newcommand{\D}{\mathbf{D}}

\newcommand{\mrm}[1]{\mathrm{#1}}

\begin{document}
\title{Experimental end-to-end demonstration of intersatellite absolute ranging for LISA}

\author{Kohei Yamamoto}
\email{y9m9k0h@gmail.com}
\affiliation{Max-Planck-Institut f\"ur Gravitationsphysik (Albert-Einstein-Institut), Callinstra\ss e 38, 30167 Hannover, Germany}
\affiliation{Leibniz Universität Hannover, Institut für Gravitationsphysik, Callinstra\ss e 38, 30167 Hannover, Germany}

\author{Iouri Bykov}
\affiliation{Max-Planck-Institut f\"ur Gravitationsphysik (Albert-Einstein-Institut), Callinstra\ss e 38, 30167 Hannover, Germany}
\affiliation{Leibniz Universität Hannover, Institut für Gravitationsphysik, Callinstra\ss e 38, 30167 Hannover, Germany}

\author{Jan Niklas Reinhardt}
\affiliation{Max-Planck-Institut f\"ur Gravitationsphysik (Albert-Einstein-Institut), Callinstra\ss e 38, 30167 Hannover, Germany}
\affiliation{Leibniz Universität Hannover, Institut für Gravitationsphysik, Callinstra\ss e 38, 30167 Hannover, Germany}

\author{Christoph Bode}
\affiliation{Max-Planck-Institut f\"ur Gravitationsphysik (Albert-Einstein-Institut), Callinstra\ss e 38, 30167 Hannover, Germany}
\affiliation{Leibniz Universität Hannover, Institut für Gravitationsphysik, Callinstra\ss e 38, 30167 Hannover, Germany}

\author{Pascal Grafe}
\affiliation{Max-Planck-Institut f\"ur Gravitationsphysik (Albert-Einstein-Institut), Callinstra\ss e 38, 30167 Hannover, Germany}
\affiliation{Leibniz Universität Hannover, Institut für Gravitationsphysik, Callinstra\ss e 38, 30167 Hannover, Germany}

\author{Martin Staab}
\affiliation{Max-Planck-Institut f\"ur Gravitationsphysik (Albert-Einstein-Institut), Callinstra\ss e 38, 30167 Hannover, Germany}
\affiliation{Leibniz Universität Hannover, Institut für Gravitationsphysik, Callinstra\ss e 38, 30167 Hannover, Germany}
\affiliation{LNE-SYRTE, Observatoire de Paris, Université PSL, CNRS, Sorbonne Université,
61 avenue de l’Observatoire, 75014 Paris, France}

\author{Narjiss Messied}
\affiliation{Max-Planck-Institut f\"ur Gravitationsphysik (Albert-Einstein-Institut), Callinstra\ss e 38, 30167 Hannover, Germany}
\affiliation{Leibniz Universität Hannover, Institut für Gravitationsphysik, Callinstra\ss e 38, 30167 Hannover, Germany}

\author{Myles Clark}
\affiliation{Max-Planck-Institut f\"ur Gravitationsphysik (Albert-Einstein-Institut), Callinstra\ss e 38, 30167 Hannover, Germany}
\affiliation{Leibniz Universität Hannover, Institut für Gravitationsphysik, Callinstra\ss e 38, 30167 Hannover, Germany}

\author{Germ\'{a}n Fern\'{a}ndez Barranco}
\affiliation{Max-Planck-Institut f\"ur Gravitationsphysik (Albert-Einstein-Institut), Callinstra\ss e 38, 30167 Hannover, Germany}
\affiliation{Leibniz Universität Hannover, Institut für Gravitationsphysik, Callinstra\ss e 38, 30167 Hannover, Germany}

\author{Thomas S. Schwarze}
\affiliation{Department of Mechanical and Aerospace Engineering, P.O. Box 116250, University of Florida, Gainesville, Florida 32611, USA}

\author{Olaf Hartwig}
\affiliation{Max-Planck-Institut f\"ur Gravitationsphysik (Albert-Einstein-Institut), Callinstra\ss e 38, 30167 Hannover, Germany}
\affiliation{Leibniz Universität Hannover, Institut für Gravitationsphysik, Callinstra\ss e 38, 30167 Hannover, Germany}
\affiliation{LNE-SYRTE, Observatoire de Paris, Université PSL, CNRS, Sorbonne Université,
61 avenue de l’Observatoire, 75014 Paris, France}
\affiliation{Max-Planck-Institut f\"ur Gravitationsphysik (Albert-Einstein-Institut), Am M\"{u}hlenberg 1, 14476 Potsdam, Germany}

\author{Juan Jos\'{e} Esteban Delgado}
\affiliation{Max-Planck-Institut f\"ur Gravitationsphysik (Albert-Einstein-Institut), Callinstra\ss e 38, 30167 Hannover, Germany}
\affiliation{Leibniz Universität Hannover, Institut für Gravitationsphysik, Callinstra\ss e 38, 30167 Hannover, Germany}

\author{Gerhard Heinzel}
\affiliation{Max-Planck-Institut f\"ur Gravitationsphysik (Albert-Einstein-Institut), Callinstra\ss e 38, 30167 Hannover, Germany}
\affiliation{Leibniz Universität Hannover, Institut für Gravitationsphysik, Callinstra\ss e 38, 30167 Hannover, Germany}

\begin{abstract}
The Laser Interferometer Space Antenna (LISA) is a gravitational wave detector in space.
It relies on a post-processing technique named time-delay interferometry (TDI) to suppress the overwhelming laser frequency noise by several orders of magnitude.
This algorithm requires intersatellite-ranging monitors to provide information on spacecraft separations.
To fulfill this requirement, we will use on-ground observatories, optical sideband-sideband beatnotes, pseudo-random noise ranging (PRNR), and time-delay interferometric ranging (TDIR).
This article reports on the experimental end-to-end demonstration of a hexagonal optical testbed used to extract absolute ranges via the optical sidebands, PRNR, and TDIR.
These were applied for clock synchronization of optical beatnote signals sampled at independent phasemeters.
We set up two possible PRNR processing schemes: Scheme 1 extracts pseudoranges from PRNR via a calibration relying on TDIR; Scheme 2 synchronizes all beatnote signals without TDIR calibration. The schemes rely on newly implemented monitors of local PRNR biases. After the necessary PRNR treatments (unwrapping, ambiguity resolution, bias correction, in-band jitter reduction, and/or calibration), Scheme 1 and 2 achieved ranging accuracies of \SI{2.0}{\centi\meter} to \SI{8.1}{\centi\meter} and \SI{5.8}{\centi\meter} to \SI{41.1}{\centi\meter}, respectively, below the classical \SI{1}{\meter} mark with margins.
\end{abstract}

\maketitle

\section{Introduction}\label{sec:introduction}
The Laser Interferometer Space Antenna (LISA) is a large-class mission by the European Space Agency (ESA) to detect gravitational waves (GWs) in space. The observation band will be shifted from the high-frequency region (\SI{10}{\Hz} to \SI{1}{\kilo\Hz}), where ground-based GW detectors~\cite{LVK2018} operate, to the \si{\milli\Hz} regime~\cite{PrePhaseA}, which is rich in GWs from systems with heavier astronomical bodies~\cite{Seoane2023}.

The LISA constellation consists of three spacecraft (SC), forming a nearly equilateral triangle with the intersatellite separation of around \SI{2.5}{million\,kilometer}.
Each SC hosts two free-falling test masses (TMs), one at the end of each intersatellite laser link. Heterodyne interferometry, in which a beam reflected by a TM surface interferes with a local beam, tracks the longitudinal motion of the TM over a broad dynamic range. The relative fluctuation of the TM-to-TM distance must be sensed with the precision of roughly \SI{10}{\pico\meter\prtHz} for GW detection.
To achieve this stringent requirement, the intersatellite link is split into one interspacecraft interferometer and two TM interferometers, one per SC. An onboard core device, called the phasemeter (PM), extracts the GW information from phases of heterodyne interferometric beatnotes via a digital phase-locked loop (DPLL). A local clock on the corresponding SC drives the PM; therefore, the beatnote phase measurements, tracked by the DPLL, are time-stamped by the onboard clock.

LISA does not actively control the separations between each SC; hence, considering the annual orbital dynamics, the distant TMs will develop relative velocities of up to \SI{\pm 10}{\meter\per\second}, resulting in MHz Doppler shifts and relative changes in the armlength up to \SI{1}{\percent}.
LISA will adopt a heterodyne frequency range from \SI{5}{\mega\Hz} to \SI{25}{\mega\Hz} for carrier-carrier beatnotes based on laser frequency planning~\cite{Heinzel2024} in order to deal with these Doppler shifts. The significant arm-length mismatches will cause a huge coupling of laser frequency noise, dominating the uncorrected detector performance around eight orders of magnitude above the target sensitivity.

Time-delay interferometry (TDI) resolves the problem by synthesizing virtual equal-arm interferometers in post-processing, using absolute arm-length information as input.
In addition, synchronization between the three independent onboard clocks must also be performed in post-processing because TDI requires combining phases from different SCs.
Intersatellite ranging and clock-tone transfer relies on two onboard functions; pseudo-random noise ranging (PRNR)~\cite{Heinzel2011,Esteban2011,Sutton2010} and \si{\giga\Hz} clock-sidebands~\cite{Hellings2001}.
These functions monitor the combination of the intersatellite light travel times and the differential clock signals, but with different features; the sidebands have very low noise in the observation band but are missing absolute range information, whereas PRNR includes the absolute range but is noisy in the observation band.

The quantity in which the clock differences and the intersatellite light travel times are entangled is called pseudorange $R^{\tau_i}_{ij}$.
This can be formulated as the difference between the local clock time of the receiving SC at the event of reception of a beam and the local clock time of the transmitting SC at the event of emission of the beam~\cite{HartwigPhD},
\begin{align}
	R^{\tau_i}_{ij}(\tau) &= \underbrace{\tau^{\tau_i}_{i}(\tau)}_{\tau} - \tau^{\tau_i}_{j}(\tau - d_{ij}(\tau)),
	\label{eq:pseudo_range}
\end{align}
where $d_{ij}$ is the intersatellite light travel time for light received on SC\,$i$ and emitted from SC\,$j$.
The superscript $\tau_i$ indicates that the quantity is evaluated according to the local clock on SC $i$.
In addition to the mentioned two functionalities, time-delay interferometric ranging (TDIR)~\cite{Tinto2005,StaabPhD} is a complementary technique to derive the pseudoranges in post-processing.
TDIR uses TDI itself as a cost function and derives optimal pseudoranges that minimize the noise power of the TDI output.
Previous research~\cite{Reinhardt2023} suggests a framework to combine these observables to gain accurate and precise pseudorange estimates.
In this article, we strictly define the term \emph{pseudorange} by~\cref{eq:pseudo_range} that regards the SC as point masses neglecting any onboard delays.
Such additional delays to the pseudoranges in any observables are called \emph{biases}.

Previously, clock synchronization among independent PMs has been demonstrated at LISA performance levels via \si{\giga\Hz}-clock modulations and TDIR with the hexagonal optical testbed, or just ``the Hexagon", at the Albert Einstein Institute (AEI) Hannover~\cite{Schwarze2019,Yamamoto2022}.
Concerning PRNR, a series of investigations under different test environments was conducted focusing on its noise property~\cite{Esteban2009,Esteban2011,Sutton2010,Sutton2013}; however, even apart from such stochastic jitters, extracting proper absolute ranges from PRNR is not trivial.
It requires overcoming several obstacles: unwrapping the raw PRNR estimates due to code repeat cycle, resolving the ambiguity due to a finite length of the pseudo-random noise (PRN) code, correcting ranging biases due to onboard delays, and finally reducing its in-band fluctuations.
Recently, the focus of studying this technology in the context of spaceborne GW detectors is evolving towards this direction~\cite{Xie2023,Euringer2023B,YamamotoPhD}.

This article reports on the experimental end-to-end demonstration of PRNR using the Hexagon.
In the Hexagon, which does not feature long-arm distances to simulate intersatellite light travel time, the pseudoranges reduce to pure clock differences by dropping $d_{ij}$ in~\cref{eq:pseudo_range}; hence, the performance will be evaluated by applying PRNR to pure clock synchronization.
PRNR is classically formulated as a ranging technology that senses biases in addition to the targets, namely pseudoranges. The biases can be caused by onboard components, like cables, or frequency-dependent delays due to photoreceivers (PRs). Hence, the biases in PRNR measurements need to be estimated and corrected.
This study demonstrates two possible PRNR post-processing schemes applicable to LISA, which uses additional local-PRN-code tracking~\cite{Sutton2013,YamamotoPhD,Euringer2023B}.
The first PRNR post-processing scheme (Scheme 1), also investigated in~\cite{YamamotoPhD} with two PMs, aligns with the classical formulation of PRNR above: it aims to extract pseudoranges from PRNR observables with the aid of TDIR-based calibration.
On the other hand, the second PRNR post-processing scheme (Scheme 2) does not completely remove such ranging biases any longer but transforms time frames including them, whose common terms do not affect TDI performances~\cite{Reinhardt2023}.
This article provides the first experimental base for putting the concept of \emph{absolute} ranging with local PRN code tracking into reality in the context of LISA.
Also, some lessons learned for TDIR will be discussed as sidenotes: the number of injected tones to break degeneracy and post-processing compensations for flexing-filtering coupling~\cite{Bayle2019}.
Furthermore, this article announces that the testbed now has all LISA technologies on the intersatellite links and can generate sets of LISA-like signals.
Therefore, it can be used to test the LISA data analysis pipeline with data from the experiment.

\section{Experimental Setup}\label{sec:experimental_setup}

\begin{figure*}
    \centering
    \includegraphics[width=12.9cm]{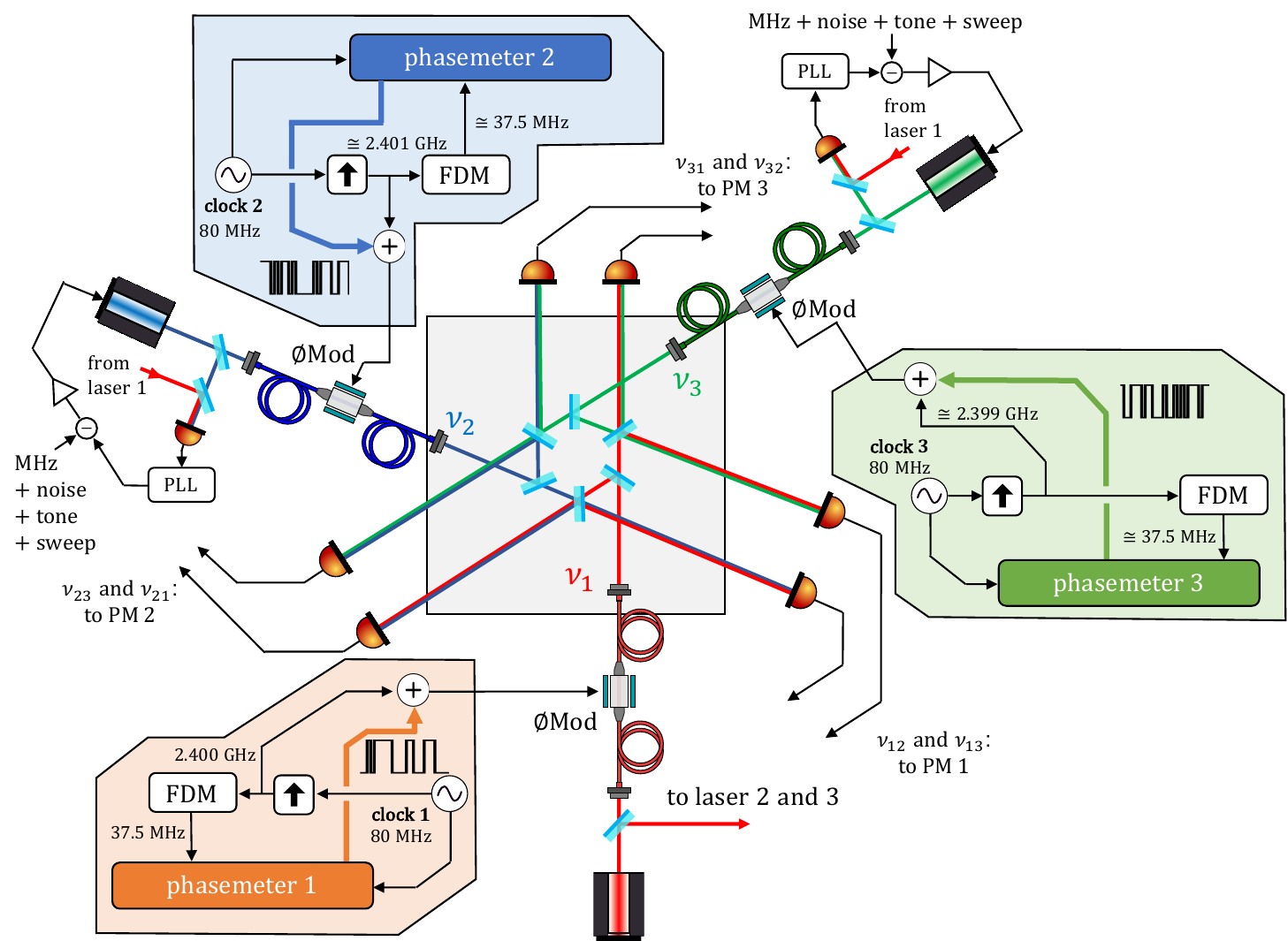}
    \caption{
        \label{fig:experimental_setup}
        A schematic of the experiment to demonstrate PRNR and clock synchronization in the Hexagon.
        The three lasers are locked to each other with \si{\mega\Hz} offsets and injections of stochastic noises, tones, and slow sweeps and are interfered pairwise to generate three optical beatnotes.
        Each module's bold and colored arrow represents the addition of a PRN-code to the \si{\giga\Hz}-clock signal.
        The modulation signals drive EOMs to phase-modulate the laser beams, denoted by ``$\phi$Mod".
        The frequency distribution module (FDM) shares the fixed division ratio 64.
    }
\end{figure*}

\cref{fig:experimental_setup} shows the experimental setup composed of three colored modules, each representing one SC in LISA.
Evolving from the previous iteration of the experiment~\cite{Yamamoto2022}, it features PRN-code modulations as drawn by a bold-colored line in each module.
The PRN code generated by the local PM is added to the \si{\giga\Hz}-clock signal with a power combiner, which then drives an electro-optic modulator (EOM).
Also, to configure bidirectional ranging links, two complementary PRs looking at different ports of a combining beam splitter are connected to different PMs. Thus, in this topology, the setup can simulate all six links of the LISA interspacecraft interferometers without long-arm entities.
Other possible experimental topologies are discussed in~\cite{YamamotoPhD}.

\begin{figure}
    \centering
    \includegraphics[width=8.6cm]{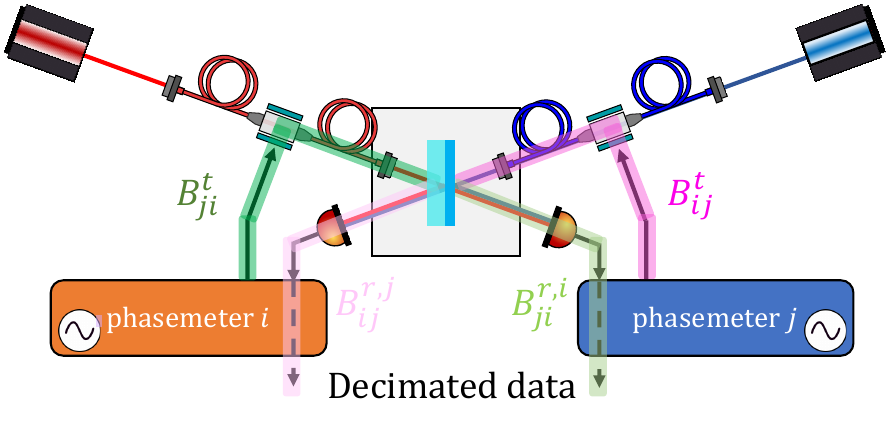}
    \caption{
        \label{fig:bidirectional_setup}
        A schematic focusing on one of the bidirectional links between two PMs.
        $B$ denotes ranging biases, split into transmitter and receiver sides, and will be discussed in \cref{sub:ranging_observables}.
    }
\end{figure}

A bidirectional optical link is featured in \cref{fig:bidirectional_setup}.
This practically corresponds to the case in LISA where the SC separation shrinks to zero and two combining beam splitters in an interspacecraft interferometer at each SC converge to one. Therefore, in the Hexagon, the events of transmission and reception occur at the same beam splitter at the same time. Biases denoted by $B_{ij}$ are split into receiver and transmitter sides, as will be elaborated in~\cref{sec:model}.

\begin{figure}
    \centering
    \includegraphics[width=8.6cm]{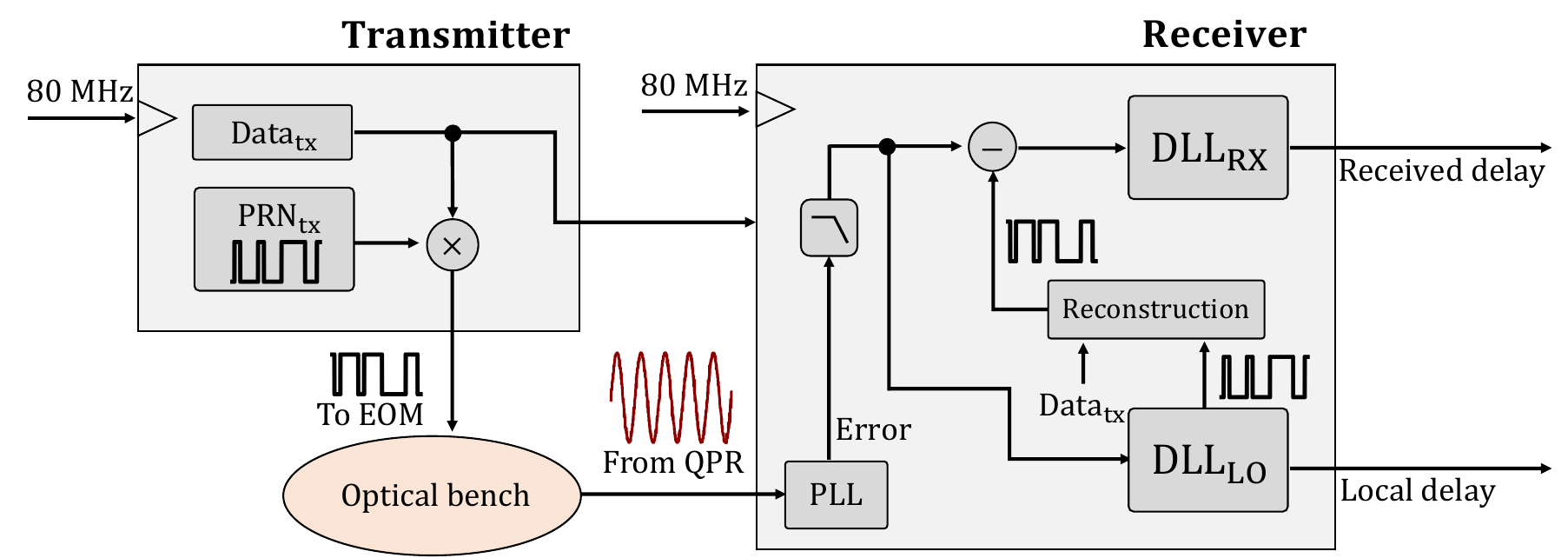}
    \caption{
        \label{fig:pm_entire_architecture}
        A simplified DLL architecture within a single PM.
        Data$_\mrm{tx}$ and PRN$_\mrm{tx}$ are the transmitted data and PRN sequence, respectively, which are superimposed before being delivered to the EOM.
        The resulting EOM phase modulation is reconstructed by DLL$_\mrm{LO}$ tracking the local code simultaneously to DLL$_\mrm{RX}$ tracking the received code. The reconstructed local code is feed-forwarded to the input of DLL$_\mrm{RX}$ for ICC.
    }
\end{figure}

In addition to modifications in the experimental setup, the PM architecture has also been developed from~\cite{Yamamoto2022}; see \cref{fig:pm_entire_architecture}.
The primary development is the implementation of a delay-locked loop (DLL) and a PRN-code generator for PRNR with a chip rate of \SI{1.25}{\mega\Hz} and a code repetition rate of around \SI{1221}{\Hz}, together with data encoding and decoding at \SI{78125}{bps}.
Two DLL instances are implemented: a receiving DLL (DLL$_\mrm{RX}$) that tracks the received PRN code and a local DLL (DLL$_\mrm{LO}$) that tracks the local code. Those DLLs acquire the target codes from the same DPLL error signal that is low-pass filtered to eliminate junks above the PRN chip rate.
As discussed in \cref{sec:model}, the local-PRNR will help monitor local biases.
In addition, DLL$_\mrm{LO}$ is also used to suppress code interference by feeding-forward the reconstructed local code to the input of DLL$_\mrm{RX}$~\cite{Sutton2013}.
This will be referred to as interfering code cancellation (ICC) for the rest of this article.
Data communication is not the scope of this article; nevertheless, random data is encoded on a transmitted PRN code at \SI{78125}{bps} because it influences code interference and PRNR.
In this case, ICC requires another internal feed-forward within the local PM: sending the encoded random data stream from the encoder to DLL$_\mrm{LO}$ to adjust its polarity.

Data decimation stages from \SI{80}{\mega Sps} to a few \si{Sps} are identical to~\cite{Yamamoto2022}: the first decimation stage with cascaded integrator-comb (CIC) filter decimates data to around \SI{610}{Sps}, which is further down-sampled via three serial decimation stages accompanied by finite impulse response (FIR) filters for the final data rate of \SI{3.4}{Sps}.~\footnote{The lastest nominal sampling rates for LISA has \SI{4}{Sps} in the end.}

To wrap up this section, it is noted that having different heterodyne frequencies at the PRs and PMs in the Hexagon is important from the perspectives of not only phase extraction tests~\cite{Schwarze2019} but also absolute ranging tests. Suppose we only have a single beatnote frequency as the bidirectional link in~\cref{fig:bidirectional_setup}. In that case, the setup cannot properly probe the impact of parameter-dependent biases (e.g., PRs) because they are common between the two signal chains. Hence, absolute-ranging investigations can only be complete with different heterodyne frequencies at different PRs, which is the case in the setup shown in~\cref{fig:experimental_setup}.

\section{Model}\label{sec:model}
This section provides the model of the optical beatnote measurements, PRNR observables, and the two PRNR processing schemes. Since clock synchronization was already formulated in~\cite{Yamamoto2022}, we omit some details here. The secondary clock $i$ can be expressed in terms of the primary clock $m$ as
\begin{align}
    \tau^{\tau_m}_i(\tau) &= \tau + \delta\tau^{\tau_m}_i(\tau)
    \nonumber\\
    &= \tau + \qt{i}(\tau) + \delta\tau_{i,0},
    \label{eq:tau_i_m}
\end{align}
where $\tau$ represents a spacecraft elapsed time (SCET) counted by the primary clock $m$, $\qt{i}$ denotes the relative timing error, and $\delta\tau_{i,0}$ is the constant initial offset between the clocks. $\delta\tau^{\tau_m}_{i}(\tau)$ is called timer deviation, which is the combination of $\qt{i}$ and $\delta\tau_{i,0}$. The superscript $\tau_m$ indicates the reference time frame.

In the following formulation, optical beatnote signals are assumed to have already been corrected by a pilot tone (PT) signal. This operation transfers the local in-band timing reference from the \SI{80}{\mega Sps} system clock to the PT signal~\cite{YamamotoPhD}. All observables modeled in this section are the low-rate PM outputs after the down-sampling, based on the assumption that decimation stages have a flat response in the observation band below \SI{1}{\Hz}.

\subsection{Beatnote signals and signal combinations}\label{sub:ifo_signals}
To start with a simple case, it is first assumed that all optical beatnote signals are recorded according to an arbitrary single primary clock $\tau_m$. This corresponds to the case where a single PM measures all signals. Also, any onboard delays, namely biases, are first neglected. A carrier-carrier beatnote frequency between beams $i$ and $j$ in the reference time frame $\tau_m$ is written by,
\begin{align}
    \nu^{\tau_m}_{c,ij}(\tau) = \nu^{\tau_m}_{c,j}(\tau) - \nu^{\tau_m}_{c,i}(\tau),
    \label{eq:nu_tau_m_ij}
\end{align}
where $\nu^{\tau_m}_{c,i}$ is the total frequency of beam $i$ evaluated in the time frame $\tau_m$. Subscripts $c$ represent carrier signals. A noise-free combination, also known as a \emph{three-signal combination}~\footnote{We discuss in \cref{sec:tdi,sec:lisa_combi} how this relates to the TDI combinations used for LISA and that this is actually the unique laser noise canceling combination for the Hexagon.}, according to the reference time frame $\tau_m$ evaluates phase extraction via the PRs and PMs,
\begin{align}
    \Delta^{\tau_{m}}_\mrm{1PM}(\tau) &= \nu^{\tau_m}_{c,23}(\tau) + \nu^{\tau_m}_{c,12}(\tau) + \nu^{\tau_m}_{c,31}(\tau) \equiv 0.
    \label{eq:Delta_1PM_tau_m}
\end{align}
It is also possible to put all six carrier signals into a combination, configuring a balanced-detection mode via a complementary pair of PRs,
\begin{align}
    \Delta^{\tau_{m}}_\mrm{1PMb}(\tau) &= \frac{1}{2}\left(\nu^{\tau_m}_{c,23}(\tau)-\nu^{\tau_m}_{c,32}(\tau)\right) + \frac{1}{2}\left(\nu^{\tau_m}_{c,12}(\tau)-\nu^{\tau_m}_{c,21}(\tau)\right) 
    \nonumber\\
    &\hspace{10mm}+ \frac{1}{2}\left(\nu^{\tau_m}_{c,31}(\tau)-\nu^{\tau_m}_{c,13}(\tau)\right) \equiv 0.
    \label{eq:Delta_1PM_tau_m_6}
\end{align}

Let's transform the single-PM case into the actual three-PM case. As mentioned, the setup in~\cref{fig:experimental_setup} configures bidirectional links for every beatnote as depicted in~\cref{fig:bidirectional_setup}. A PM\,$i$ receives and time-stamps two interferometric signals by its timing reference $\tau_i$,
\begin{align}
    \nu^{\tau_i}_{c,ij}(\tau) = \nu^{\tau_i}_{c,j}(\tau) - \nu^{\tau_i}_{c,i}(\tau),
    \label{eq:nu_tau_i_ij}\\
    \nu^{\tau_i}_{c,ik}(\tau) = \nu^{\tau_i}_{c,k}(\tau) - \nu^{\tau_i}_{c,i}(\tau).
    \label{eq:nu_tau_i_ik}
\end{align}

To make use of all carrier signals in a final combination, the locally measured carrier signals are first combined,
\begin{align}
    \nu'^{\tau_i}_{c,jk}(\tau) = \nu^{\tau_i}_{c,ik}(\tau) - \nu^{\tau_i}_{c,ij}(\tau).
    \label{eq:nu_tau_i_jk}
\end{align}
The final combination is constructed via clock synchronization of two secondary PMs\,$j$ and $k$ to a primary PM\,$i$,
\begin{align}
    \Delta^{\tau_{i}}_\mrm{3PM}(\tau;\delta\hat\tau_{j,0},\delta\hat\tau_{k,0}) &= \frac{1}{2}\left(\nu'^{\tau_i}_{c,jk}(\tau) + \tilde{\nu'}^{\tau_j}_{c,ki}(\tau) + \tilde{\nu'}^{\tau_k}_{c,ij}(\tau)\right)
    \nonumber\\
    &\approx \Delta^{\tau_{i}}_\mrm{1PMb}(\tau),
    \label{eq:Delta_3PM_tau_i}
\end{align}
where $\tilde{\nu}^{\tau_j}$ is a frequency signal synchronized to the reference frame $\tau_i$ via a differential clock signal between the PM\,$i$ and PM\,$j$ at the PM\,$i$ and an initial timer offset $\delta\hat\tau_{j,0}$. The initial offsets $\delta\hat\tau_{j,0}$ and $\delta\hat\tau_{k,0}$ can be estimated using TDIR, as demonstrated in~\cite{Yamamoto2022}, or PRNR, as described in the following sections.

To wrap up this section, we formulate carrier-carrier beatnote signals with biases, which were omitted for conciseness in the formulation above. Only the receiver sides, namely biases after the combining beam splitters, are important because of the symmetrical hexagonal bench. The beatnote signal can be rewritten with such a beatnote receiver timestamp bias $B^r_{c,ij}$:
\begin{align}
    \nu^{\tau_i}_{c,ij}(\tau) &\rightarrow \nu^{\tau_i}_{c,ij}\left(\tau - B^r_{c,ij}(\tau)\right)
    \nonumber\\
    &\rightarrow \nu^{\tau_i}_{c,ij}\left(\tau - B^{r}_{c,ij,\mrm{BS\rightarrow DPLL}}(\tau)\right)
    \label{eq:nu_tau_i_ij_delay}\\
    B^r_{c,ij}(\tau) &= B^{r}_{c,ij,\mrm{BS\rightarrow DPLL}}(\tau) + B^{r}_{c,\mrm{DPLL\rightarrow DEC}},
    \label{eq:B_c_ij}
\end{align}
where $B^r_{c,ij}$ was split into two parts. $B^{r}_{c,ij,\mrm{BS\rightarrow DPLL}}$ is the bias from the combining beam splitter to the input of the DPLL, including the parameter-dependent contribution, like the transfer functions of the PRs and the PM analog electronics. On the other hand, $B^{r}_{c,\mrm{DPLL\rightarrow DEC}}$ is the bias from the DPLL to the decimated data.
$B^{r}_{c,\mrm{DPLL\rightarrow DEC}}$ will not be actively considered in the following sections because all beatnotes are expected to share the same biases here to the accuracy we are interested in with ranging; hence, they do not couple to the final performance. The PRNR processing schemes in~\cref{sub:prnr_scheme1,sub:prnr_scheme2} will be formulated as if $B^{r}_{c,ij,\mrm{BS\rightarrow DPLL}}$ was the only beatnote bias based on the second row in \cref{eq:nu_tau_i_ij_delay}.

\subsection{PRNR observables}\label{sub:ranging_observables}
PRNR involves an ambiguity $L_\mrm{code}$ in the order of \SI{100}{\kilo\meter} due to the finite code length: the code repetition rate $f_\mrm{code}$ is, as mentioned in \cref{sec:experimental_setup}, \SI{1221}{\Hz}, which results in the ambiguity of around \SI{246}{\kilo\meter} via $L_\mrm{code} = c/f_\mrm{code}$ where $c$ is a speed of light.
The associated ambiguity offset $a_\mrm{prn}$ can be expressed as
\begin{align}
	a_\mrm{prn} &= \text{round}\left[\delta\hat\tau_{i,0}/(L_\mrm{code}/c)\right] \cdot (L_\mrm{code}/c).
	\label{eq:prn_amb_offset}
\end{align}
A separate ranging monitor is necessary to resolve the ambiguity $L_\mrm{code}$. TDIR is available in the Hexagon experiment. In LISA, we additionally have ground-based observations, which comprise orbit determinations and MOC time correlations~\cite{Reinhardt2023}. Below, we assume the PRNR ambiguity $L_\mrm{code}$ to be already resolved.

Pseudoranges in the Hexagon are equivalent to differential clock offsets because the experiment does not simulate interspacecraft light travel times. Hence, any physical signal travel time is absorbed into a ranging bias $B_{ij}$ (see \cref{fig:bidirectional_setup}). The PRNR estimate $R^{\mrm{prn},\tau_i}_{ij}(\tau)$ is given by,
\begin{align}
	R^{\mrm{prn},\tau_i}_{ij}(\tau) &= \tau^{\tau_i}_{i}(\tau) - \tau^{\tau_i}_{j}(\tau - B_{ij}(\tau)),
	\label{eq:mpr_ij_tau_m}
\end{align}
which corresponds to the replacement of $d_{ij}$ in \cref{eq:pseudo_range} with $B_{ij}$. The superscript ``prn" is used to clearly distinguish the PRNR estimate from the TDIR counterpart, which will be formulated later. Notice that the ranging bias $B_{ij}$ was assumed time-variant. In-band stochastic ranging noises are omitted in this formulation.

\cref{fig:bidirectional_setup} decomposed the ranging bias into the transmitter and receiver sides,
\begin{align}
	B_{ij}(\tau) = B^t_{ij} + B^{r,j}_{ij}(\tau),
	\label{eq:B_ij_decomp}
\end{align}
where the superscripts $t$ and $r$ represent the transmitter and receiver biases, respectively. Notice that the time dependency is only applied to the receiver bias. The transmitter side contains digital signal processing, electronics, cables, and stable optics (fibers, an EOM, optical paths on an optical bench, etc.). Hence, it wouldn't drift significantly against the scale that PRNR probes: the EOM could cause thermally induced drifts of a few centimeters~\cite{Barke2009,BarkePhD} or \SI{10}{\meter} silica optical fibers could result in the length change of \OSI{1}{\milli\meter} over thermal drifts by \SI{10}{degree}. On the other hand, some receiver biases are caused by more complex mechanisms because PRN signals go through the system with frequency-dependent and parameter-dependent group delays, e.g., the PRs and DPLLs. This could yield non-negligible time dependencies of the receiver bias. In addition, the receiver bias slightly depends on the specific PRN sequence~\cite{Euringer2023A}; hence, the superscript $j$ was added to indicate the SC at which the code was created.

Applying the timer model in~\cref{eq:tau_i_m}, the received- and local-PRNR observables at a PM\,$i$ with a link to a PM\,$j$ can be formulated as follows,
\begin{align}
	R^{\mrm{prn},\tau_i}_{ij}(\tau) &= \tau^{\tau_i}_{i}(\tau) - \tau^{\tau_i}_{j}(\tau - B_{ij}(\tau))
	\nonumber\\
	&= B_{ij}(\tau) - \delta\tau^{\tau_i}_j(\tau - B_{ij}(\tau)) \hspace{5mm} \text{for DLL$_\mrm{RX}$},
	\label{eq:mpr_ij_tau_i}\\
	R^{\mrm{prn},\tau_i}_{ii,j}(\tau) & = \tau^{\tau_i}_{i}(\tau) - \tau^{\tau_i}_{i}(\tau - B_{ii,j}(\tau))
	\nonumber\\
	&= B_{ii,j}(\tau) \hspace{29mm} \text{for DLL$_\mrm{LO}$},
	\label{eq:mpr_iij_tau_i}
\end{align}
where the total bias in the local PRNR is defined by,
\begin{align}
	B_{ii,j}(\tau) = B^t_{ji} + B^{r,i}_{ij}(\tau).
	\label{eq:B_ii_decomp}
\end{align}
The index notation was slightly extended for the local-PRNR observable $B_{ii,j}$: the subscript $ii$ indicates that the PRN code originates and terminates on the same ``SC" or PM\,$i$, while the subscript $j$ refers to the SC it faces. The receiver biases in local and received PRNR ($B^{r,i}_{ij}$ in \cref{eq:B_ii_decomp} and $B^{r,j}_{ij}$ in \cref{eq:B_ij_decomp}) slightly differ due to non-identical code sequences, even though their signal chains are identical.

Finally, the ranging estimate via a sideband-sideband beatnote and TDIR is defined as a ranging reference of the experiment,
\begin{align}
	R^{\mrm{tdir+sb},\tau_i}_{ij}(\tau) &= - \qt{j}(\tau - B_{ij}(\tau)) - \delta\hat\tau^\mrm{tdir}_{j,0}
	\nonumber\\
	&\approx - \delta\tau^{\tau_i}_j(\tau - B_{ij}(\tau))
    \nonumber\\
    &\approx - \delta\tau^{\tau_i}_j(\tau)
	\label{eq:tdir_ij_tau_i}
\end{align}
where $\qt{j}$ is the accumulated sideband measurement and $\delta\hat\tau^\mrm{tdir}_{j,0}$ is the initial offset fitted via TDIR. The third row is valid if the cross term between $B_{ij}$ and $\qdt{j}$ is below the target accuracy $\sim$\OSI{1}{\nano\second}, which is normally the case. Non-common receiver biases among carrier-carrier beatnotes parameterize $\delta\hat\tau^\mrm{tdir}_{j,0}$. Therefore, the change of heterodyne frequencies couples to TDIR estimates $\delta\hat\tau^\mrm{tdir}_{j,0}$ via non-constant group delays of components like the PRs, if not treated. Once such contributors are removed, TDIR (+ sideband measurements) acts as a pseudorange reference of the experiment, namely $R^{\mrm{tdir+sb},\tau_i}_{ij}\approx- \delta\tau^{\tau_i}_j(\tau)$. The trip of a clock signal from PM\,$j$ to PM\,$i$ considers the same bias $B_{ij}$ as the corresponding PRN signal $R^{\mrm{prn},\tau_i}_{ij}$, which is strictly not the case.  Nevertheless, $B_{ij}$ is used here in the same logic as the third row: neglecting the first-order term of such a bias difference between the \si{\giga\Hz}-clock sideband and the PRN signal in Taylor expansion, based on the assumption that its product with a sub-ppm differential clock drift $\qdt{j}$ is negligible.

\subsection{PRNR processing scheme 1}\label{sub:prnr_scheme1}

This article extends Scheme 1, already presented in~\cite{YamamotoPhD} with two PMs, to three independent PMs. The aim of Scheme 1 is to extract pseudoranges from the PRNR observables in~\cref{eq:mpr_ij_tau_i,eq:mpr_iij_tau_i}.

The experiment evaluates the absolute-ranging feature of PRNR via comparison with TDIR (+ sideband measurements). For example, the received-PRNR estimate in~\cref{eq:mpr_ij_tau_i} can be compared with the TDIR estimate in \cref{eq:tdir_ij_tau_i} to reveal its bias contribution $B_{ij}$,
\begin{align}
	R^{\mrm{prn},\tau_i}_{ij}(\tau) &\approx B_{ij}(\tau) + R^{\mrm{tdir+sb},\tau_i}_{ij}(\tau).
	\label{eq:prn_rx_vs_tdir}
\end{align}

The bias in the received PRNR can be mostly removed via the local PRNR,
\begin{align}
	R^{\mrm{prn},\tau_i}_{ij, \mrm{corr}}(\tau) &= R^{\mrm{prn},\tau_i}_{ij}(\tau)-R^{\mrm{prn},\tau_i}_{ii,j}(\tau)
	\nonumber\\
	&\approx R^{\mrm{tdir+sb},\tau_i}_{ij}(\tau) + \Delta B_{ij}(\tau),
	\label{eq:prn_corr_vs_tdir}\\
	\Delta B_{ij}(\tau) &= \Delta B^t_{ij} + \Delta B^r_{ij}(\tau),
	\label{eq:prn_corr_bias}\\
	\Delta B^t_{ij} &= B^t_{ij} - B^t_{ji},
	\label{eq:Delta_trans_bias}\\
	\Delta B^r_{ij}(\tau) &= B^{r,j}_{ij}(\tau) - B^{r,i}_{ij}(\tau).
	\label{eq:Delta_receive_bias}
\end{align}
The residual bias $\Delta B_{ij}$ has two components: $\Delta B^t_{ij}$, referred to as a transmitter bias mismatch, and $\Delta B^r_{ij}(\tau)$, referred to as a receiver bias residual. Remarkably, the time dependency of the receiver bias can be highly suppressed because the two codes go through the identical signal path on the receiver side. Scheme 1 requires prior calibration to derive the residual bias $\Delta B_{ij}$. Then, in science mode, the bias is expected to be further suppressed with the accuracy of the residual time dependency $\delta B^r_{ij}$,
\begin{align}
	R^{\mrm{prn},\tau_i}_{ij, \mrm{cal}}(\tau) &= R^{\mrm{prn},\tau_i}_{ij, \mrm{corr}}(\tau) - \Delta B_{ij,\:\rm{cal}}
    \nonumber\\
    &= R^{\mrm{prn},\tau_i}_{ij, \mrm{corr}}(\tau) - \left(\Delta B^t_{ij} + \Delta B^r_{ij,\:\rm{cal}}\right)
	\nonumber\\
	&= - \delta\tau^{\tau_i}_j(\tau - B_{ij}(\tau)) + \delta B^r_{ij}(\tau).
	\label{eq:prn_cal_vs_tdir}\\
	\delta B^r_{ij}(\tau) &= \Delta B^r_{ij}(\tau) - \Delta B^r_{ij,\:\rm{cal}},
	\label{eq:delta_receive_bias}
\end{align}
where $\Delta B_{ij,\:\rm{cal}}$ is the total residual bias calibrated in prior calibration, which is the sum of the transmitter bias mismatch $\Delta B^t_{ij}$ and the receiver bias residual in prior calibration $\Delta B^r_{ij,\:\rm{cal}}$.

Finally, PRNR is applied to clock synchronization. We normally rely on clock sideband measurements to precisely monitor the in-band stochastic jitter, while PRNR accurately monitors large time offsets. To combine those two monitors to acquire precise and accurate ranging estimates, this article follows the averaging method presented in~\cite{Hartwig2022}: the initial offset $\delta\tau_{i,0}$ is derived by the averaged difference between the PRNR estimate and the integration of the sideband measurement $\qt{j}$,
\begin{align}
    R^{\mrm{prn},\mrm{sb},\tau_i}_{ij,x}(\tau) &= \qt{j}(\tau - B_{ij}(\tau)) + \delta\hat\tau^\mrm{prn}_{j,0},
    \label{eq:prnr_timer_deviation}\\
    \delta\hat\tau^\mrm{prn}_{j,0} &= \mrm{avg}[R^{\mrm{prn},\tau_i}_{ij,x}(\tau)-\qt{j}(\tau - B_{ij}(\tau))].
    \label{eq:prnr_timer_offset}
\end{align}
This strongly suppresses PRNR in-band stochastic jitters via the number of averaged samples.\footnote{Alternatively, PRNR and sideband measurements can be combined in a Kalman-like filter as proposed in \cite{Reinhardt2023}. This opens up the possibility to include proper noise models for variables and measurements.} The in-band jitter in the local PRNR observables is mitigated by applying a polynomial fit or a low-pass filter in post-processing, omitted in the formulation above. If the residual jitter is negligible against $\delta B^r_{ij}(\tau)$, the PRNR performance in clock synchronization would be expected to be limited by the ranging bias estimation.

We remark that non-common beatnote biases, most importantly time-dependent contributors, need to be also corrected to keep the TDIR-calibrated biases $\Delta B_{ij,\:\rm{cal}}$ valid based on \cref{eq:tdir_ij_tau_i} for long-term; otherwise, beatnote biases would drift from the values with which TDIR calibration was conducted, and $\delta B^r_{ij}$ in~\cref{eq:delta_receive_bias} amounts to a non-negligible scale.
For this purpose, the receiver signal-chain transfer functions can also be calibrated separately, and the time dependency can be removed by advancing beatnote signals in \cref{eq:nu_tau_i_ij_delay},
\begin{align}
    \A^{\tau_i}_{c,ij}\nu^{\tau_i}_{c,ij}\left(\tau - B^{r}_{c,ij,\mrm{BS\rightarrow DPLL}}(\tau)\right) &= \nu^{\tau_i}_{c,ij}\left(\tau\right).
    \label{eq:Acij_nucij}
\end{align}
where $\A^{\tau_i}_{c,ij}$ is an advancement operator for a beatnote signal by a group delay at the frequency $\nu^{\tau_i}_{c,ij}$.

In summary, Scheme 1 aims to extract pseudoranges from the received PRNR measurements by means of the local PRNR measurements and TDIR-based calibration. To apply the pseudoranges for clock synchronization, the beatnote signals must be advanced based on the signal chain transfer functions as shown in~\cref{eq:Acij_nucij}. This logic aligns with the bias correction procedure in~\cite{Reinhardt2023}, where onboard biases are removed separately from PRNR and beatnote signals at the first data analysis stage.

\subsection{PRNR processing scheme 2}\label{sub:prnr_scheme2}
Scheme 2 does not extract pseudoranges any longer. Instead, it focuses on the proper input to TDI when no advancements are applied to the beatnotes. The previous study~\cite{Reinhardt2023} clarified this as the difference between the SCETs associated with the appearance of any instantaneous feature of laser noise in the beatnote phase measurements on the local and the remote SC. Scheme 1 applies the beatnote advancement (see~\cref{eq:Acij_nucij}) to align the pseudoranges with this proper TDI input. However, following the description above, TDI would not necessarily require the pseudoranges (in this article's definition); hence, Scheme 2 focuses on the needs of TDI and tries to meet it most efficiently with PRNR and no further reference measurements (TDIR + sideband measurements).

Scheme 1 corrected the beatnote bias $B^{r}_{c,ij,\mrm{BS\rightarrow DPLL}}$ via the calibrated signal-chain transfer function. Contrary to Scheme 1, Scheme 2 will correct it via PRNR measurements. Hence, we remove the potential biggest mismatch between the beatnote and PRN signal chains: the DPLLs and pre-DLL low-pass filters (see \cref{fig:pm_entire_architecture}). The DPLL transfer functions cause such a mismatch because of the difference in readout points. We first subtract the sum of those contributions $B^{r,i}_{ij,\mrm{DPLL\rightarrow LPF}}$ from any PRNR observable via numerical simulations,

\begin{align}
    R^{\mrm{prn},\tau_i}_{ij,\mrm{\rightarrow DPLL}}(\tau) &= R^{\mrm{prn},\tau_i}_{ij}(\tau) - B^{r,j}_{ij,\mrm{DPLL\rightarrow LPF}}(\tau)
    \nonumber\\
    &= B_{ij}(\tau) - B^{r,j}_{ij,\mrm{DPLL\rightarrow LPF}}(\tau) - \delta\tau^{\tau_i}_j\left(\tau - B_{ij}(\tau)\right)
    \nonumber\\
    &\approx B_{ij,\mrm{\rightarrow DPLL}}(\tau) - \delta\tau^{\tau_i}_j\left(\tau - B_{ij,\mrm{\rightarrow DPLL}}(\tau)\right)
    \label{eq:Rprn_DPLL_LPF_ij_tau_i}\\
    R^{\mrm{prn},\tau_i}_{ii,j,\mrm{\rightarrow DPLL}}(\tau) &= R^{\mrm{prn},\tau_i}_{ii,j}(\tau) - B^{r,i}_{ij,\mrm{DPLL\rightarrow LPF}}(\tau)
    \nonumber\\
    &= B_{ii,j}(\tau) - B^{r,i}_{ij,\mrm{DPLL\rightarrow LPF}}(\tau)
    \nonumber\\
    &= B_{ii,j,\mrm{\rightarrow DPLL}}(\tau)
    \label{eq:Rprn_DPLL_LPF_iij_tau_i}
\end{align}
since
\begin{align}
    B_{ij,\mrm{\rightarrow DPLL}}(\tau) &\coloneqq B_{ij}(\tau) - B^{r,j}_{ij,\mrm{DPLL\rightarrow LPF}}(\tau)
    \nonumber\\
    &= B^t_{ij} + B^{r,j}_{ij,\mrm{BS\rightarrow DPLL}}(\tau)
    \label{eq:Bij_to_DPLL}\\
    B_{ii,j,\mrm{\rightarrow DPLL}}(\tau) &\coloneqq B_{ii,j}(\tau) - B^{r,i}_{ij,\mrm{DPLL\rightarrow LPF}}(\tau)
    \nonumber\\
    &= B^t_{ji} + B^{r,i}_{ij,\mrm{BS\rightarrow DPLL}}(\tau)
    \label{eq:Biij_to_DPLL}
\end{align}
where $B^{r,j}_{ij,\mrm{BS\rightarrow DPLL}}$ and $B^{r,i}_{ij,\mrm{BS\rightarrow DPLL}}$ are, in analogy to $B^{r}_{c,ij,\mrm{BS\rightarrow DPLL}}$ for beatnote signals, the biases from the combining beam splitter to the input of the DPLL for the PRN signals. Notice that PRNR biases after the pre-DLL low-pass filter are not considered in \cref{eq:Rprn_DPLL_LPF_ij_tau_i,eq:Rprn_DPLL_LPF_iij_tau_i},  because they are expected to be common in all PRNR observables to the target accuracy $\sim$\SI{1}{\nano\second} or \SI{1}{\meter} in range, i.e., the same rationale as $B^{r}_{c,\mrm{DPLL\rightarrow DEC}}$ for the beatnote signals.

Second, delay and advancement operators based on $R^{\mrm{prn},\tau_i}_{ij,\mrm{\rightarrow DPLL}}$ and $R^{\mrm{prn},\tau_i}_{ii,j,\mrm{\rightarrow DPLL}}$ are defined,
\begin{align}
    \A^{\mrm{prn},\tau_a}_b x(\tau) &= x\left(\tau + R^{\mrm{prn},\tau_a}_{b,\mrm{\rightarrow DPLL}}(\tau)\right),
    \label{eq:Aprn}\\
    \D^{\mrm{prn},\tau_a}_b x(\tau) &= x\left(\tau - R^{\mrm{prn},\tau_a}_{b,\mrm{\rightarrow DPLL}}(\tau)\right),
    \label{eq:Dprn}
\end{align}
where $a$ and $b$ are the arbitrary expressions of clock indices and their combinations. For example, if $a=i$ and $b=ij$ (or $b=ii,j$), \cref{eq:Aprn,eq:Dprn} represent the operators based on the received (or local) PRNR observable at a PM\,$i$ with a link to a PM\,$j$.
The beatnote signal can be advanced by the corresponding operator of the local-PRNR observable as,
\begin{align}
    \A^{\mrm{prn},\tau_i}_{ii,j}&\nu^{\tau_i}_{c,ij}\left(\tau - B^{r}_{c,ij,\mrm{BS\rightarrow DPLL}}(\tau)\right)
    \nonumber\\
    &= \nu^{\tau_i}_{c,ij}\left(\tau + B^t_{ji} + \delta B^{r,i}_{ij,\mrm{BS\rightarrow DPLL}}(\tau)\right)
    \nonumber\\
    &\approx \nu^{\tau_i}_{c,ij}\left(\tau + B^t_{ji}\right),
    \label{eq:nu_ij_A}\\
    \delta B^{r,i}_{ij,\mrm{BS\rightarrow DPLL}}(\tau) &\coloneqq B^{r,i}_{ij,\mrm{BS\rightarrow DPLL}}(\tau) - B^{r}_{c,ij,\mrm{BS\rightarrow DPLL}}(\tau),
    \label{eq:delta_Br_c_ij}
\end{align}
where $\delta B^{r,i}_{ij,\mrm{BS\rightarrow DPLL}}$ is the mismatch in the bias from the combining beamsplitter to the input of the DPLL between the beatnote and PRN signals. This would be expected to be relatively small because the PRN power spectrum is spread symmetrically around the carrier-carrier beatnote~\cite{Euringer2023B}; therefore, it was dropped in \cref{eq:nu_ij_A}. However, it should be remarked that this depends on the specifications of the electronics on the chains and is the potential limiting factor of the performance of Scheme 2.

Third, to transform the time frame from clock $i$ to $j$, the delay operator based on the received PRNR measurement at a PM\,$j$ with a link to a PM\,$i$ can be applied,
\begin{widetext}
    \begin{align}
    \D^{\mrm{prn},\tau_j}_{ji}\A^{\mrm{prn},\tau_i}_{ii,j}\nu^{\tau_i}_{c,ij}\left(\tau - B^{r}_{c,ij,\mrm{BS\rightarrow DPLL}}(\tau)\right) &\approx \D^{\mrm{prn},\tau_j}_{ji}\nu^{\tau_i}_{c,ij}\left(\tau + B^t_{ji}\right)
        \nonumber\\
        &\approx \nu^{\tau_i}_{c,ij}\left(\tau - B^{r,i}_{ji,\mrm{BS\rightarrow DPLL}}(\tau)+\delta\tau^{\tau_j}_i\left(\tau - B^{r,i}_{ji,\mrm{BS\rightarrow DPLL}}(\tau)\right)\right)
        \nonumber\\
        &=\nu^{\tau_i}_{c,ij}\left(\tau^{\tau_j}_i\left(\tau - B^{r,i}_{ji,\mrm{BS\rightarrow DPLL}}(\tau)\right)\right)
        \nonumber\\
        &= \nu^{\tau_j}_{c,ij}\left(\tau - B^{r,i}_{ji,\mrm{BS\rightarrow DPLL}}(\tau)\right)
        \nonumber\\
        &\approx \nu^{\tau_j}_{c,ij}\left(\tau - B^{r}_{c,ji,\mrm{BS\rightarrow DPLL}}(\tau)\right),
        \label{eq:nu_ij_DA}
    \end{align}
\end{widetext}
where the transmitter bias $B^t_{ji}$ is canceled between the advancement and delay operators. The third line applies the basic formula of the timer model in~\cref{eq:tau_i_m}. The fifth line neglects the bias mismatch between the interferometric signal and the PRNR observable $\delta B^{r,i}_{ji,\mrm{BS\rightarrow DPLL}}$. Also, the Doppler factor, in general, required for clock synchronization of beatnote frequencies is omitted for simplicity. \cref{eq:nu_ij_DA} is the essence of clock synchronization with PRNR in this article. This scheme does not correct the bias at PM\,$j$, namely $B^{r}_{c,ji,\mrm{BS\rightarrow DPLL}}$; however, it is not an issue because the reference signal $\nu^{\tau_j}_{c,ji}$ also share the same receiver bias, that is, this bias is common between the two beatnote signals. Therefore, \cref{{eq:nu_ij_DA}} properly time-transforms $\nu^{\tau_i}_{c,ij}$ to $\nu^{\tau_j}_{c,ji}$ without the extraction of the pseudoranges from the PRNR measurements.

\begin{figure}
    \centering
    \includegraphics[width=8.6cm]{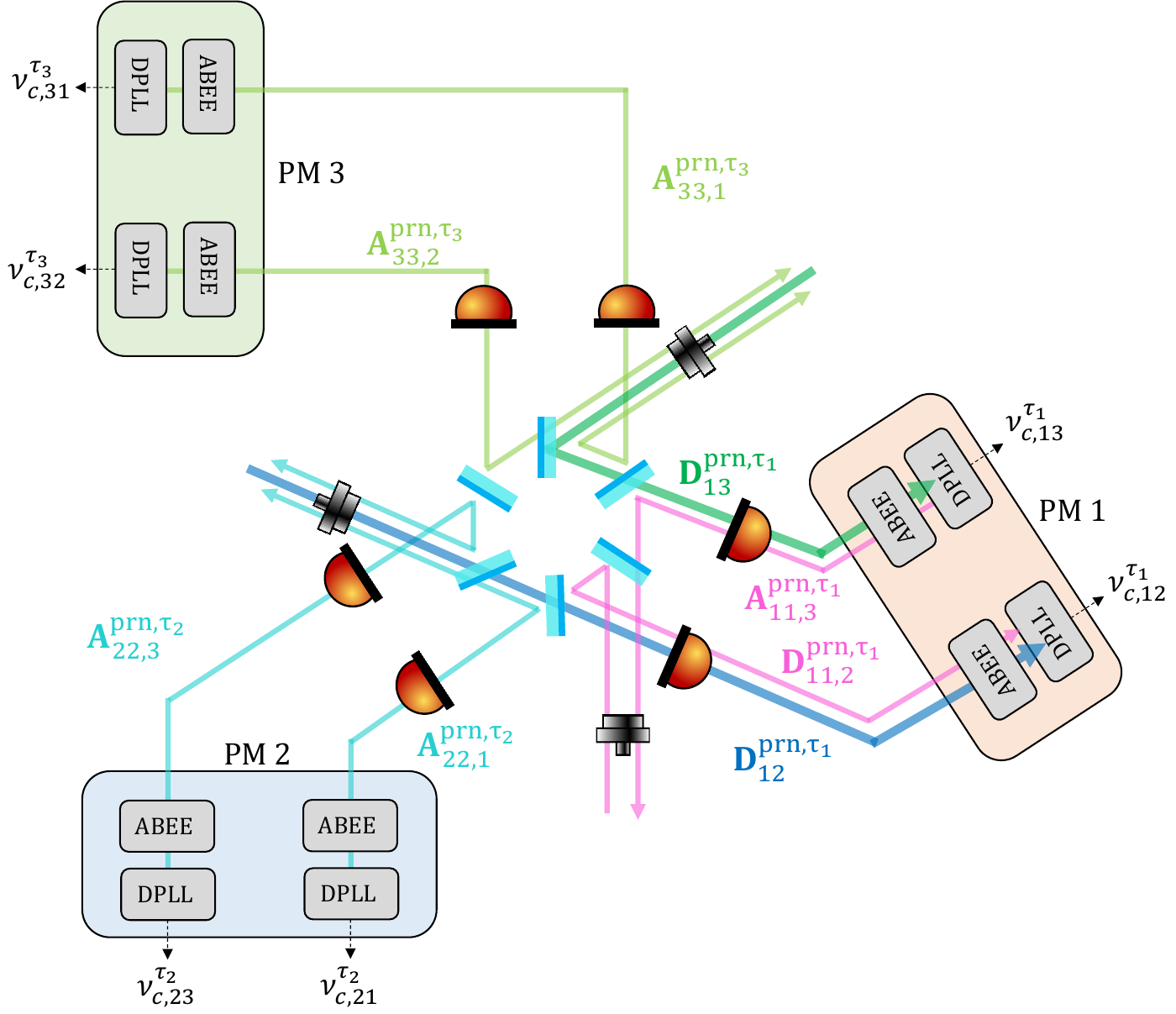}
    \caption{
        \label{fig:new_prnr_scheme}
        A diagram of the PRNR protocol without bias correction. In this example, the interferometric signals at PM 2 and 3 and $\nu^{\tau_1}_{c,13}$ at PM 1 are shifted to the readout time of $\nu^{\tau_1}_{c,12}$ at PM 1. ABEE stands for analog back-end electronics.
    }
\end{figure}

Finally, the entire PRNR-based clock synchronization for the final signal combination in \cref{eq:Delta_3PM_tau_i} is provided. \cref{fig:new_prnr_scheme} is the diagram visually depicting how the interferometric signals $\nu^{\tau_i}_{c,ij}$ are time-transformed to the reference signal $\nu^{\tau_1}_{c,12}$ over the setup with those PRNR-based advancement and delay operators. The corresponding signal combination can be written by,
\begin{widetext}
    \begin{align}
        2\Delta^{\tau_{1}}_\mrm{3PM} &= \nu^{\tau_1}_{c,12} + \D^{\mrm{prn},\tau_1}_{12}\left(\A^{\mrm{prn},\tau_2}_{22,1}\nu^{\tau_2}_{c,21} + \A^{\mrm{prn},\tau_2}_{22,3}\nu^{\tau_2}_{c,23}\right)
        + \D^{\mrm{prn},\tau_1}_{11,2}\A^{\mrm{prn},\tau_1}_{11,3}\Bigg\{\Bigg.
        \nu^{\tau_1}_{c,13} + \D^{\mrm{prn},\tau_1}_{13}\left(\A^{\mrm{prn},\tau_3}_{33,1}\nu^{\tau_3}_{c,31} + \A^{\mrm{prn},\tau_3}_{33,2}\nu^{\tau_3}_{c,32}\right)\Bigg.\Bigg\}
        \nonumber\\
        &\approx \nu^{\tau_1}_{c,12} + \D^{\mrm{prn},\tau_1}_{11,2}\A^{\mrm{prn},\tau_1}_{11,3}\nu^{\tau_1}_{c,13} 
        \nonumber\\
        &\hspace{8mm} + \D^{\mrm{prn},\tau_1}_{12}\left(\A^{\mrm{prn},\tau_2}_{22,1}\nu^{\tau_2}_{c,21} + \A^{\mrm{prn},\tau_2}_{22,3}\nu^{\tau_2}_{c,23}\right)
        \nonumber\\
        &\hspace{8mm} + \D^{\mrm{prn},\tau_1}_{13}\D^{\mrm{prn},\tau_1}_{11,2}\A^{\mrm{prn},\tau_1}_{11,3}\left(\A^{\mrm{prn},\tau_3}_{33,1}\nu^{\tau_3}_{c,31} + \A^{\mrm{prn},\tau_3}_{33,2}\nu^{\tau_3}_{c,32}\right),
        \label{eq:Delta_3PM_tau_1_prnr}
    \end{align}
\end{widetext}
where the time arguments of beatnote signals and the Doppler factors are omitted for simplicity. Notice that this combination synchronizes not only the signals at PM\,2 and\,3 to the linked signals at PM\,1 but also the local time difference at PM\,1 using $\D^{\mrm{prn},\tau_1}_{11,3}\A^{\mrm{prn},\tau_1}_{11,3}$. The second line exchanges the order of $\D^{\mrm{prn},\tau_1}_{11,3}\A^{\mrm{prn},\tau_1}_{11,3}$ and $\D^{\mrm{prn},\tau_1}_{13}$~\footnote{They are not completely commutative because of the ramp of $\D^{\mrm{prn},\tau_1}_{13}$ due to the clock frequency offsets $\sim$\OSI{0.1}{ppm}; however, its cross term with $\D^{\mrm{prn},\tau_1}_{11,3}\A^{\mrm{prn},\tau_1}_{11,3}$ $\sim$\OSI{10}{\meter} (or equivalently $\sim$\OSI{100}{\nano\second}) results in $\sim$\OSI{10}{\femto\second} and is negligible against the scale PRNR looks at, i.e., \OSI{1}{\nano\second}.}, which would ease data analysis by applying the timing transformation between PM 1 and PM3 at the end. The same combination can be configured for any reference PM by cyclically shifting the PM indices.

The formulation above did not consider the PRNR in-band jitter suppression. In actual data processing, the same as Scheme 1, \cref{eq:prnr_timer_deviation,eq:prnr_timer_offset} and the low-pass filtering are applied to mitigate the significant in-band jitter of the received and local PRNR measurements, respectively. After that, the signal combination in~\cref{eq:Delta_3PM_tau_1_prnr} performs.

\section{Results}\label{sec:results}

The PRNR processing schemes modelled in \cref{sec:model} were demonstrated with the Hexagon experiment depicted in \cref{fig:experimental_setup}. After providing general information about the test environment, key information for evaluating the two PRNR processing schemes will be described in~\cref{sub:testbed_capability}. \cref{sub:demo_prnr_1,sub:demo_prnr_2} report on experimental demonstrations of Scheme 1 and 2, respectively.

For ranging tests, controlling laser frequency noise plays a pivotal role. Therefore, some dynamics were intentionally injected to beatnote frequencies via laser lock loops: a white frequency noise at \SI{60}{\Hz\prtHz} to simulate the LISA-like beatnote noise level, in-band sinusoidal tones to ease the evaluation of ranging performance, and slow sweeps to simulate Doppler effects over the LISA heterodyne bandwidth (but with much faster speeds than real Doppler effects in LISA to fit them in the timescale of a lab experiment). The slow sweeps are important to demonstrate its coupling to the ranging biases due to non-constant group delays caused by components on the beatnote signal chains after the combining beam splitters. The measured bandwidths of the signal chains were within $\SI{38.5}{}\pm\SI{5.0}{\mega\Hz}$.

\begin{figure}
    \centering
    \includegraphics[width=8.6cm]{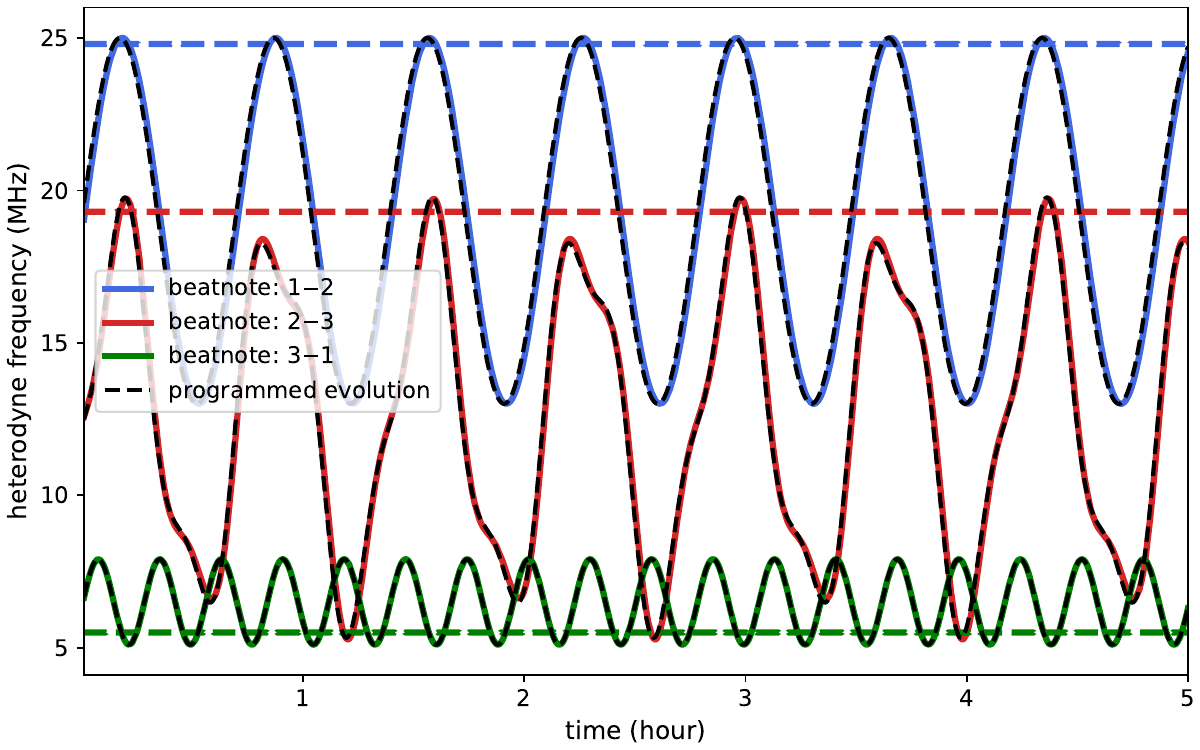}
    \caption{\label{fig:het_freq_drift_t} Injected drifts of heterodyne frequencies via laser lock. The first \SI{5}{hours} are shown out of \SI{23000}{\second} and \SI{70000}{\second} for prior calibration and science mode, respectively. Solid-colored: measured heterodyne frequencies in science mode. Dashed-colored: measured heterodyne frequencies in prior calibration. Dashed-black: programmed evolution.}
\end{figure}

Two measurements were conducted: with (\emph{science mode}) and without (\emph{calibration mode}) frequency sweeping. We first acquired calibration offsets via TDIR in calibration mode with static beatnote frequencies. After that, we conducted the primary measurement in science mode and investigated the ranging performances under frequency sweeping. The data from the two measurement modes will be analyzed based on Scheme 1 in \cref{sub:demo_prnr_1} and Scheme 2 in \cref{sub:demo_prnr_2}. Calibration mode utilized heterodyne frequencies of \SI{24.8}{\mega\Hz}, \SI{19.3}{\mega\Hz}, and \SI{5.5}{\mega\Hz} for $\nu_{c,12}$, $\nu_{c,23}$, and $\nu_{c,31}$, respectively, while the frequencies of lasers 2 and 3 were sinusoidally modulated in science mode,
\begin{align}
    O_{12}(\tau) &= \SI{6}{\mega\Hz}\cdot \sin\left(2\pi\cdot\SI{0.4}{\milli\Hz}\cdot\tau\right) + \SI{19}{\mega\Hz},
    \label{eq:O_12_sci}\\
    O_{13}(\tau) &= \SI{1.4}{\mega\Hz}\cdot \sin\left(2\pi\cdot\SI{1.0}{\milli\Hz}\cdot\tau\right) + \SI{6.5}{\mega\Hz},
    \label{eq:O_13_sci}
\end{align}
where $O_{ij}$ is a \si{\mega\Hz} offset frequency between beams $i$ and $j$. \cref{fig:het_freq_drift_t} shows the heterodyne frequencies over the first \SI{5}{hours} out of the entire measurements, which lasted for \SI{23000}{\second} and \SI{70000}{\second} in calibration and science modes, respectively.

In every measurement, all beams share the power of \OSI{100}{\micro\watt} at the PRs. The PRN modulation depth was set to \SI{0.1}{\radian}, which represents \SI{1}{\percent} of the total beam power, and random data sequences were encoded on each PRN-code stream. The \si{\giga\Hz}-clock modulation used a modulation index of \SI{0.4}{\radian}, and an offset frequency equal to that of the PRN chip rate (\SI{1.25}{\mega\Hz}) to mitigate the impact of the PRN power on the sideband readout. Concerning the time-shifting of measured beatnote frequencies, we used Lagrange interpolation with order 121 as implemented in PyTDI~\cite{pytdi}.

\subsection{Testbed capability}\label{sub:testbed_capability}

The testbed capability is demonstrated here using data taken in calibration mode. \cref{tab:key_params} summarizes the key parameters of the Hexagon.

The in-band sinusoidal modulation of the beatnote frequencies will be used to evaluate PRNR performance; however, we can also use it to maximize TDIR accuracy. This is called \emph{tone-assisted TDIR}~\cite{Mitryk2012,Francis2015}. This helps acquire good calibration offsets for PRNR in science mode. The three-signal combination requires two tones to break the degeneracy between the two timer deviations. Therefore, one tone was injected per one secondary laser. The tones were selected to have an amplitude of \SI{1}{\kilo\Hz} and frequencies of \SI{0.6226}{\Hz} and \SI{0.7620}{\Hz} for lasers 3 and 2 (i.e., $\nu_{T3}$ and $\nu_{T2}$ introduced in \cref{sec:tone_tdir}), respectively. See \cref{sec:tone_tdir} for a detailed description of this technique in the Hexagon.

\begin{figure}
    \centering
    \includegraphics[width=8.6cm]{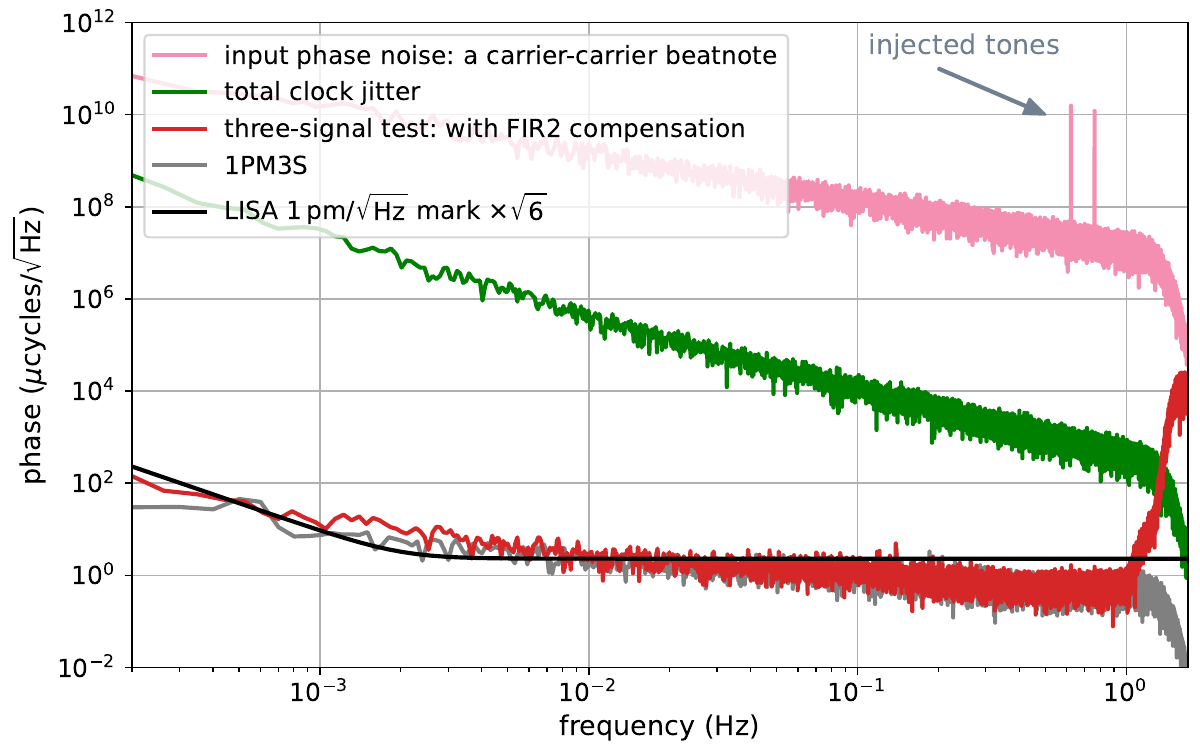}
    \caption{\label{fig:tone_tdir_f} Clock synchronization among three independent PMs via tone-assisted TDIR with static heterodyne frequencies (i.e., \SI{24.8}{\mega\Hz}, \SI{19.3}{\mega\Hz}, and \SI{5.5}{\mega\Hz}) in prior calibration. Pink: input carrier-carrier beatnote phase with two injected tones. The injected white noise at \SI{60}{\Hz\prtHz} gets the $1/f$ shape via unit conversion from \si{\Hz} to \si{cycle}. Green: total in-band clock jitter. Red: three-phasemeter three-signal combination $\Delta^{\tau_{1}}_\mrm{3PM}$ from \cref{eq:Delta_3PM_tau_i} with compensation for the second FIR filter; also see \cref{sec:tone_tdir}. Grey: three-signal combination with a single PM, i.e., $\Delta^{\tau_{1}}_\mrm{1PMb}$ in \cref{eq:Delta_1PM_tau_m_6}, giving the testbed sensitivity for clock synchronization. Black: \SI{1}{\pico\meter\prtHz} mark, considering the incoherent sum of PM channels for a requirement on phase extraction.}
\end{figure}

\cref{fig:tone_tdir_f} shows the clock synchronization performance with tone-assisted TDIR in calibration mode. The residual tones in the final signal combination in red were suppressed down to the stochastic secondary noise floor in grey, which was \SI{0.550}{\micro{cycles}\prtHz} (separately computed by Logarithmic frequency axis Power Spectral Density (LPSD)~\cite{TROBS2006}) around the tone frequencies. The secondary noise floor (i.e., the testbed sensitivity) was separately measured by connecting all six signals to a single PM, which corresponds to $\Delta^{\tau_{1}}_\mrm{1PMb}$ in~\cref{eq:Delta_1PM_tau_m_6}. This noise floor enables tone-assisted TDIR to reach millimeter accuracy over an averaging time of \OSI{10000}{\second}.
Concerning the clock jitter in green, the measured in-band noise in this experiment can be modeled by $10^{-11}/f^{1.5}$\,\si{\second\prtHz}~\cite{YamamotoPhD}.
This performance is much worse over the observation band than the expected space-qualified clock stability $10^{-14}/f^{0.5}$\,\si{\second\prtHz}~\cite{Hartwig2022}; hence, the experiment performs as the worst-case demonstration in this regard.

The performance in~\cref{fig:tone_tdir_f}, namely below \SI{1}{\pico\meter\prtHz} at high frequencies,  confirms that PRN codes used for this experiment are well designed from the perspective of DPLL. PRN modulations can generally introduce noise that impacts carrier (and sideband) phase tracking. To mitigate this, custom PRN codes were designed to attenuate their effect. The PRN-code-induced additive noise for carrier phase extraction is expected to be around \SI{10}{\nano\radian\prtHz}, or equivalently around \SI{1}{\femto\meter\prtHz}, at \SI{1}{\Hz} and proportionally decrease toward low frequencies~\cite{YamamotoPhD}. As mentioned, adjusting the sideband offset frequency mitigated the same PRN additive noise for sideband phase extraction.

Alongside phase measurements and TDIR, ICC aims to subtract an interfering code utilizing DLL$_\mrm{LO}$, as described in \cref{sec:experimental_setup}, to minimize an induced noise power on the DLL$_\mrm{RX}$ readout. Hence, the impact of code interference on the received PRNR is expected to be somewhat suppressed.
Distortions in the PRN signals limit the effectiveness of ICC, as these distortions introduce residual noise after subtracting the ideal, undistorted PRN replica.
Notice that ICC cannot be implemented the other way around to suppress DLL$_\mrm{LO}$ in-band noise due to the necessity to know what data was superimposed onto the PRN sequences in order to subtract them. Therefore, DLL$_\mrm{LO}$ is expected to be always nosier than DLL$_\mrm{RX}$.

\begin{table}[b]
    \caption{\label{tab:key_params}
    Key parameters of the testbed. $T_\mrm{avg}$ is an averaging time. PRNR noise powers are computed with the average to a bandwidth of a Nyquist frequency, i.e., half of $\SI{3.391}{Sps}$. Only the carrier phase noise due to PRN modulations is estimated numerically. $c$ is a speed of light.}
    \begin{ruledtabular}
    \begin{tabular}{ll}
    Parameter & Value\\
    \hline
    testbed sensitivity around tones & \SI{0.550}{\micro{cycles}\prtHz}
    \\
    carrier phase noise due to PRN mod. & \SI{10}{\nano\radian\prtHz} at \SI{1}{\Hz}
    \\
    tone-assisted TDIR accuracy & $\SI{2.037e-9}{}\cdot c/\sqrt{T_\mrm{avg}}\hspace{1mm}\text{m}$
    \\
    receiver signal-chain bandwidth & $\SI{38.5}{}\pm\SI{5.0}{\mega\Hz}$
    \\
    
    PRNR noise power: & \\
    \hspace{10mm}$\text{PM}2\rightarrow \text{PM}1$: (RX, LO) & (\SI{0.196}{}, \SI{0.617}{}) \si{\meter\,rms}\\
    \hspace{10mm}$\text{PM}3\rightarrow \text{PM}1$: (RX, LO) & (\SI{0.189}{}, \SI{0.751}{}) \si{\meter\,rms}\\
    \hspace{10mm}$\text{PM}3\rightarrow \text{PM}2$: (RX, LO) & (\SI{0.081}{}, \SI{0.668}{}) \si{\meter\,rms}\\
    \hspace{10mm}$\text{PM}1\rightarrow \text{PM}2$: (RX, LO) & (\SI{0.148}{}, \SI{0.697}{}) \si{\meter\,rms}\\
    \hspace{10mm}$\text{PM}1\rightarrow \text{PM}3$: (RX, LO) & (\SI{0.178}{}, \SI{0.601}{}) \si{\meter\,rms}\\
    \hspace{10mm}$\text{PM}2\rightarrow \text{PM}3$: (RX, LO) & (\SI{0.143}{}, \SI{0.508}{}) \si{\meter\,rms}\\
    \end{tabular}
    \end{ruledtabular}
\end{table}

\subsection{Demonstration of PRNR Scheme 1}\label{sub:demo_prnr_1}

\begin{figure*}
    \centering
    \includegraphics[width=17.2cm]{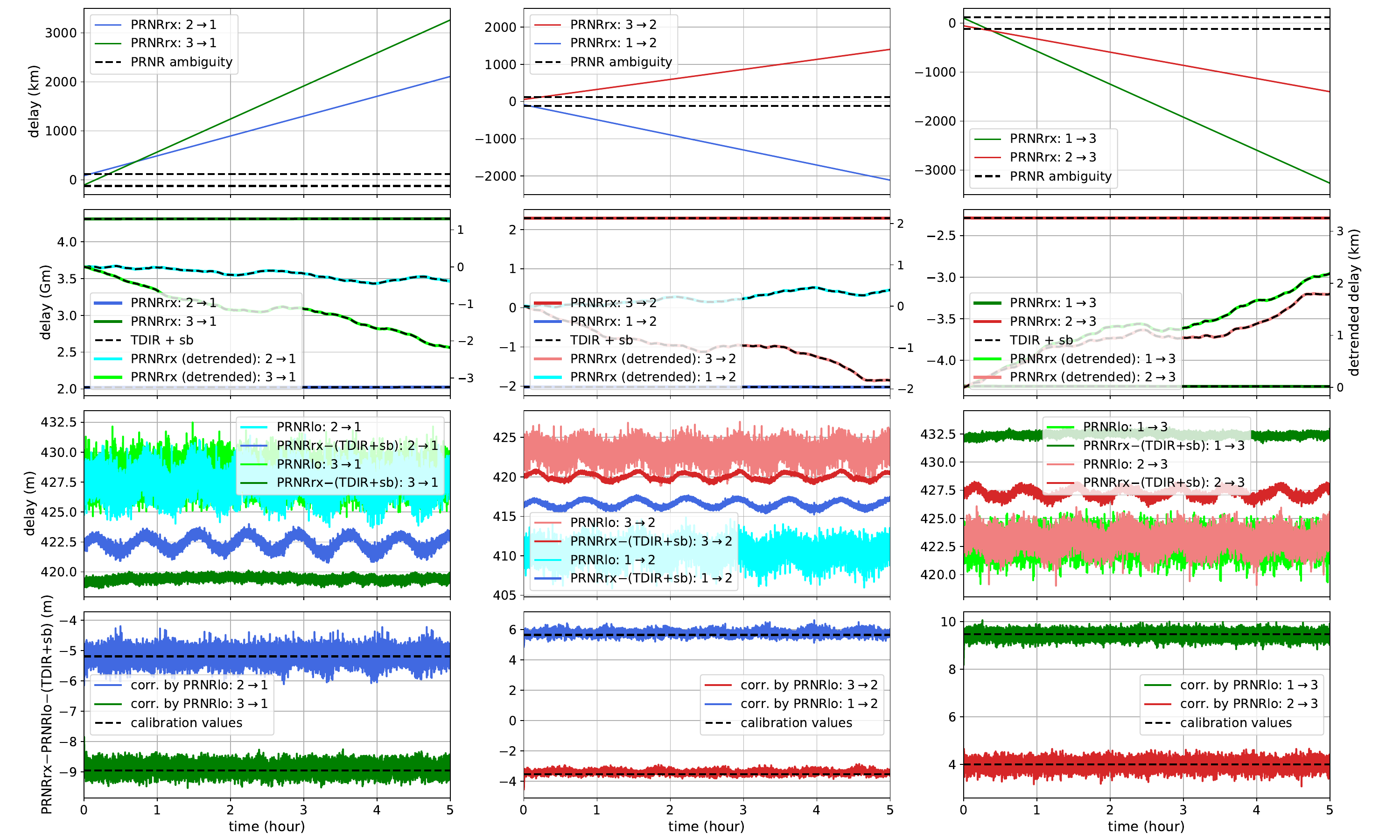}
    \caption{\label{fig:prnr_entire_process_t_1} PRNR observables, treatments, and comparisons to sideband measurements and TDIR at different PMs in science mode in Scheme 1: PM\,1 (left), PM\,2 (middle), PM\,3 (right). Only the first \SI{5}{hours} from the entire \SI{70000}{\second} measurements are shown for visibility. First row: unwrapped receive-PRNR estimates initially confined in PRNR ambiguity. Second row: ambiguity-resolved PRNR estimates via TDIR and comparisons to TDIR and sidebands (dashed-black). The total timer deviations are plotted in darker colors attached to the left y-axis, while the stochastic components after detrending are plotted in lighter colors attached to the right. Third row: differences between the received-PRNR estimates and the combination of TDIR and sidebands in the second row, revealing biases. The local-PRNR estimates are also plotted in lighter colors. Fourth row: comparison of the residual bias after the correction via local PRNR with calibration values from prior calibration (dashed-black).}
\end{figure*}

Scheme 1 modelled in~\cref{sub:prnr_scheme1} was demonstrated using the measurements in calibration and science modes. Apart from ranging observables, an important treatment in this mode is the correction of parameter-dependent beatnote biases in~\cref{eq:Acij_nucij}. This way, the pseudoranges (i.e., the pure clock differences in the Hexagon) become the essential quantities for clock synchronization via PRNR.
Otherwise, the calibration offsets acquired by TDIR in calibration mode are no longer accurate for science mode because the two modes do not share the same heterodyne frequencies, as shown in~\cref{fig:het_freq_drift_t}.

The entire PRNR treatments in science mode (ambiguity resolution, bias correction, and calibration) are summarized in \cref{fig:prnr_entire_process_t_1}: at PM\,1 (left), PM\,2 (middle), and PM\,3 (right). The top panels show the unwrapped received PRNR estimates. They are initially confined in the PRNR ambiguity (dashed-black) and then evolve almost linearly due to clock frequency offsets of around \SI{0.2}{ppm} to \SI{0.6}{ppm}. Wrapping, or jumps in data due to code repeat cycle, caused artifacts in the decimation filters, which resulted in issues during the unwrapping step in post-processing. This was omitted in the plot, as this issue was solved using a preliminary solution: DLL$_\mrm{RX}$ generates two delays that are shifted by half of the PRN code length~\cite{YamamotoPhD}.

Ambiguity resolution requires an independent ranging monitor. As mentioned in~\cref{sub:ranging_observables}, only TDIR is available in this experiment; hence, we also performed TDIR in science mode for this purpose~\footnote{Because the PRNR ambiguity is about \SI{246}{\kilo\meter} in this experiment, normal TDIR without tones, which can reach an accuracy of a few meters, is enough; nevertheless, tone-assisted TDIR was performed as a performance reference.}. The second panels from the top in~\cref{fig:prnr_entire_process_t_1} showed the ambiguity-resolved received PRNR measurements. The resulting PRNR estimates are plotted in dark colors, including huge initial offsets ($\sim$\OSI{1}{\giga\meter}, or equivalently a few seconds). In addition, by detrending those total PRNR estimates, the stochastic components of differential timer deviations are revealed (light-colored with the right y-axis). They are confirmed to be consistent with the sideband signals in dashed black; hence, the in-band noise reduction properly functions by combining the received PRNR measurements with sideband measurements, following \cref{eq:prnr_timer_deviation,eq:prnr_timer_offset}.


In the third panels from the top, to reveal the biases in the received-PRNR estimates, the differences between PRNR and TDIR (+ sideband measurements) are computed, which corresponds to $R^{\mrm{prn},\tau_i}_{ij}-R^{\mrm{tdir+sb},\tau_i}_{ij} \approx B_{ij}$ from \cref{eq:prn_rx_vs_tdir}. The frequency modulations in \cref{fig:het_freq_drift_t} coupled to the biases as expected. The local-PRNR estimates $R^{\mrm{prn},\tau_i}_{ii,j}$ are also plotted in the same axes with light colors. Notice that the local estimates include higher fluctuations due to the lack of ICC. Most of the overall bias, approximately a few hundred meters, comes from digital signal processing on the FPGA.

Finally, the bottom panels show the bias correction of the received PRNR estimates with the local ones, smoothed by a \SI{10}{\milli\Hz} low-pass filter. They confirm that biases can be mostly corrected via the local PRNR, leaving constant residuals of a few meters dominated by the transmitter bias mismatch $\Delta B^t_{ij}$ in~\cref{eq:Delta_trans_bias}.
Also, the parameter-dependent components of PRNR biases induced by heterodyne frequency drifts were highly attenuated via correction from dozens of centimeters to a few centimeters, as summarized in \cref{tab:resi_tones_1}, which meets the \SI{1}{\meter} requirement of inter-satellite ranging~\cite{LISARedBook}. This attenuation of the parameter dependency enhances the robustness of PRNR. The coupling mechanism via the DPLLs was confirmed by numerical simulation as a dominant source of the parameter dependency~\cite{YamamotoPhD}. This point will be revisited in~\cref{sub:demo_prnr_2}. The modulation correction performance at PM\,2 was limited to around \SI{7}{\centi\meter}, remarkably worse than the other measurements and the separate simulation results. The source of this limitation remains to be investigated. The residual biases after the correction in the plots agree with calibration values in dashed-black. This agreement also reflects removing parameter-dependent biases in beatnote signals, as mentioned at the beginning of this section. Subtracting the calibration values can further improve the PRNR performance, following \cref{eq:prn_cal_vs_tdir}.

\begin{figure}
    \centering
    \includegraphics[width=8.6cm]{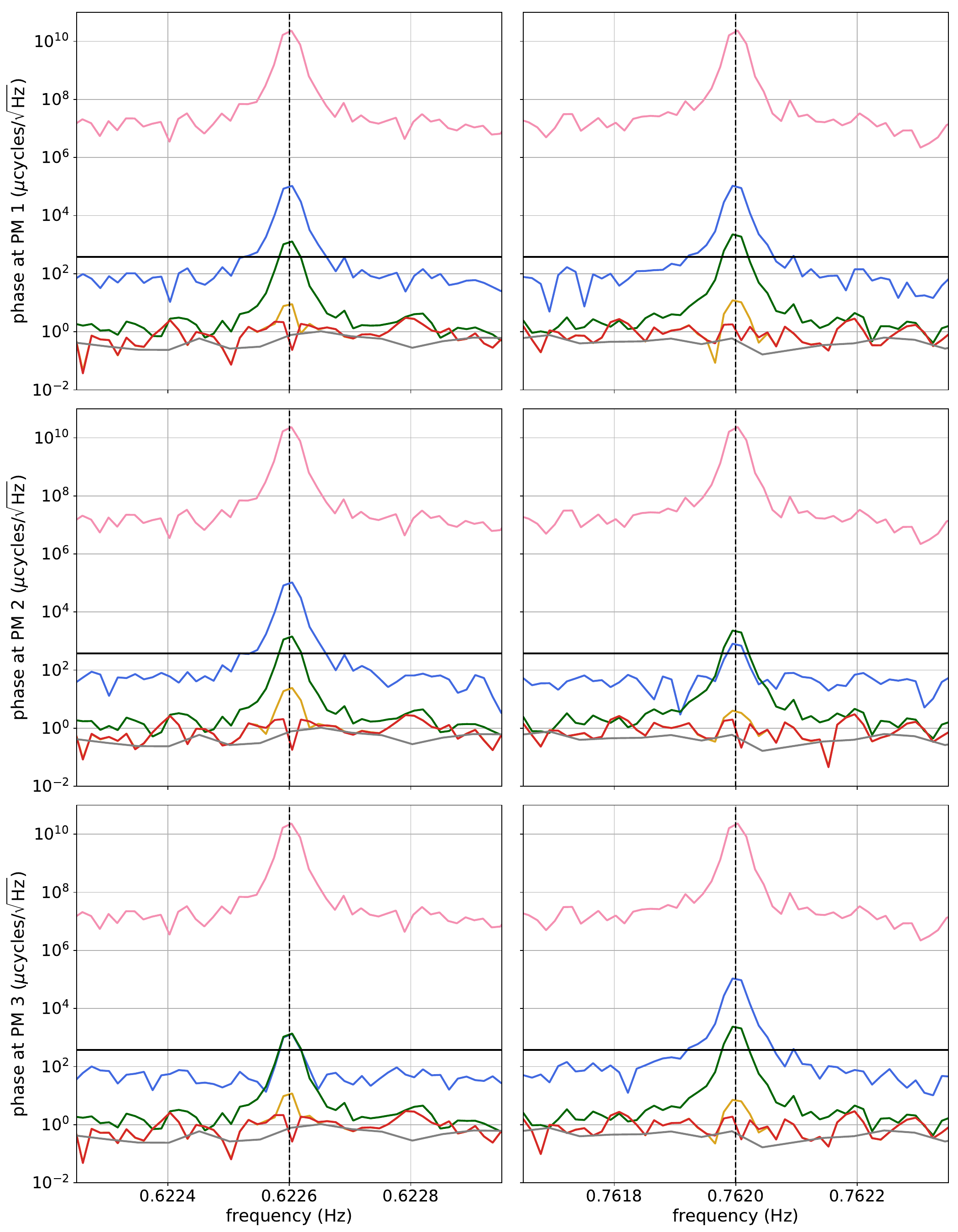}
    \caption{\label{fig:prnr_sync_f_1} Clock synchronization $\Delta^{\tau_{i}}_\mrm{3PM}$ from \cref{eq:Delta_3PM_tau_i} using the PRNR estimates at different stages in the entire treatment. From the top row, phase tones in amplitude spectral density are shown in the reference time frames of PMs 1, 2, and 3. The left and right panels show plots around tones at \SI{0.6226}{\Hz} and \SI{0.7620}{\Hz} (denoted by the dashed-black vertical lines), respectively. Pink: input carrier-carrier beatnote phase. Blue, green, gold, and red: three-signal combinations with the three PMs $\Delta^{\tau_{i}}_\mrm{3PM}$ with ambiguity-resolved PRNR $R^{\mrm{prn},\tau_i}_{ij}$ in~\cref{eq:mpr_ij_tau_i}, bias-corrected PRNR $R^{\mrm{prn},\tau_i}_{ij, \mrm{corr}}$ in~\cref{eq:prn_corr_vs_tdir}, calibrated PRNR $R^{\mrm{prn},\tau_i}_{ij, \mrm{cal}}$ in~\cref{eq:prn_cal_vs_tdir}, and tone-assisted TDIR with compensation for the second FIR filter, respectively. Grey: three-signal combination with a single PM $\Delta^{\tau_{1}}_\mrm{1PMb}$, identical to the one in \cref{fig:tone_tdir_f}. Solid-black: \SI{1}{\meter} mark for residual tones, derived by applying the corresponding delay operator to the injected tone in pink.}
\end{figure}

After suppressing in-band PRNR jitter according to \cref{eq:prnr_timer_deviation,eq:prnr_timer_offset}, PRNR was lastly applied to clock synchronization at each PM. \cref{fig:prnr_sync_f_1} shows the results at PM\,1 (top), PM\,2 (middle), and PM\,3 (bottom). The residual tones were improved as PRNR bias treatments were refined. The calibrated-PRNR estimates (gold) successfully suppressed the residual tones below the classical \SI{1}{\meter} mark (i.e., the black horizontal line), leaving \SI{1}{\centi\meter} to \SI{8}{\centi\meter} residuals. Such ranging errors derived by comparing the residual tones with the injected ones are also summarized in \cref{tab:resi_tones_1}. The results also show the significance of compensating for the parameter-dependent biases of the beatnote signals to let the pseudoranges, free from heterodyne frequencies, become the target of any ranging processing in this scheme. The combination of the residual PRNR-bias modulations and the modelling error of the transfer functions from the PRs to the DPLLs limit the final performances.

Note that \cref{fig:prnr_sync_f_1} also confirms the coupling of the ranging errors into the tones, as shown in \cref{eq:3s_tau_1,eq:3s_tau_2,eq:3s_tau_3}. In some cases (i.e., the middle right and the bottom left), clock synchronization with the ambiguity-resolved PRNR (blue) has already reached the accuracy of a few meters, even without bias correction via the local PRNR. This is because the differences between two timer deviations couple to the tones, namely $\dot{\nu}^{\tau_2}_{T2}(\tau)\cdot\left(\delta\tau^{\tau_2}_{3,\mrm{err}} - \delta\tau^{\tau_2}_{1,\mrm{err}}\right)$ and $\dot{\nu}^{\tau_3}_{T3}(\tau)\cdot\left(\delta\tau^{\tau_3}_{1,\mrm{err}} - \delta\tau^{\tau_3}_{2,\mrm{err}}\right)$ in~\cref{eq:3s_tau_2,eq:3s_tau_3}, respectively. Therefore, the biases of a few hundred meters shown in~\cref{fig:prnr_entire_process_t_1} cancel between the two received PRNR estimates, which leaves residuals of a few meters.

\begin{table*}[hbt!]
    \caption{\label{tab:resi_tones_1}
    The results of absolute ranging in the three-signal combination at each PM in Scheme 1: two links per the primary PM (opposing PMs configuring particular links are described below the primary). Ranging performances in clock synchronization, which are the main results, were computed from the residual tones in \cref{fig:prnr_sync_f_1}. Each tone residual was transformed to each pair of PMs according to the coupling in \cref{eq:3s_tau_1,eq:3s_tau_2,eq:3s_tau_3}. The heterodyne frequency drifts, shown in \cref{fig:het_freq_drift_t}, cause the PRNR-bias and beatnote-delay modulations.}
    \begin{ruledtabular}
    \begin{tabular}{lcccccc}
    Parameter &\multicolumn{2}{c}{At PM\,1} &\multicolumn{2}{c}{At PM\,2} &\multicolumn{2}{c}{At PM\,3}
    \\
     &PM\,2 &PM\,3 &PM\,3 &PM\,1 &PM\,1 &PM\,2
    \\
    \hline
    Ranging performance in clock synchronization (\si{\centi\meter}): & \\
    \hspace{5mm} With compensation for beatnote biases & \SI{3.03}{} & \SI{3.27}{} & \SI{8.14}{} & \SI{8.07}{} & \SI{1.95}{} & \SI{4.39}{} \\
    \hspace{5mm} Without compensation for beatnote biases & \SI{38.8}{} & \SI{10.1}{} & \SI{28.2}{} & \SI{27.9}{} & \SI{8.73}{} & \SI{40.7}{} \\

    Measured PRNR-bias modulation peak (\si{\centi\meter}): & & & & & & \\
    \hspace{5mm} With correction by the local PRNR & \SI{1.88}{} & \SI{1.12}{} & \SI{6.98}{} & \SI{7.25}{} & \SI{0.61}{} & \SI{1.83}{} \\
    \hspace{5mm} Without correction by the local PRNR & \SI{71.1}{} & \SI{6.99}{} & \SI{63.9}{} & \SI{60.4}{} & \SI{7.83}{} & \SI{55.6}{} \\

    Modelled beatnote-delay modulation peak (\si{\centi\meter}) & \SI{14.3}{} & \SI{1.48}{} & \SI{22.0}{} & \SI{12.3}{} & \SI{1.26}{} & \SI{13.1}{}\\
    
    \end{tabular}
    \end{ruledtabular}
\end{table*}

\subsection{Demonstration of PRNR Scheme 2}\label{sub:demo_prnr_2}

\begin{figure}
    \centering
    \includegraphics[width=8.6cm]{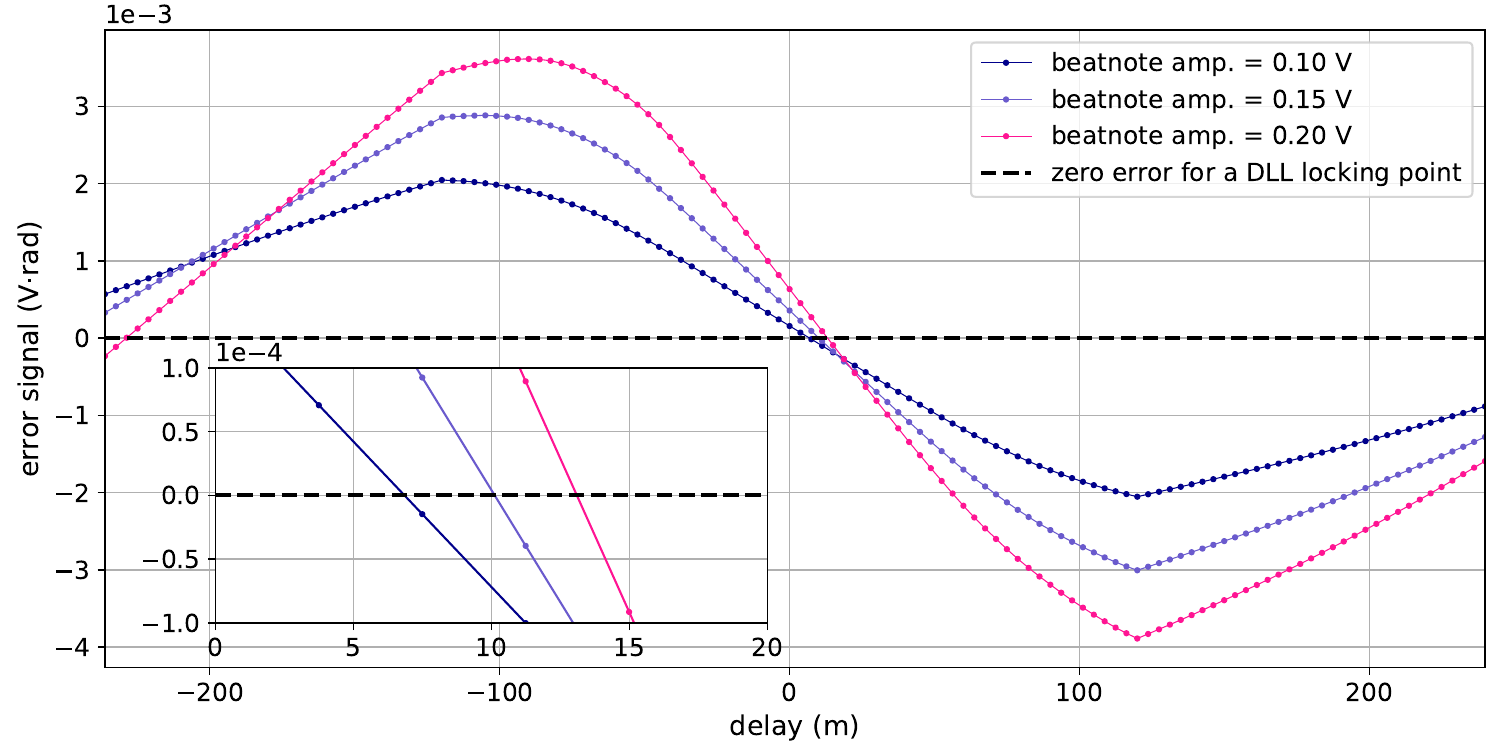}
    \caption{\label{fig:simulated_dll_error_signal_bitbalanced} Simulated DLL error signal with the input PRN signal distorted by the DPLL transfer function. The DPLL transfer function depends on the input beatnote amplitude. The small axis is a zoom-in plot around the locking point.}
\end{figure}

Scheme 2 in \cref{sub:prnr_scheme2} was also demonstrated using the measurement in science mode. We first highlight the important initial treatment in \cref{eq:Rprn_DPLL_LPF_ij_tau_i,eq:Rprn_DPLL_LPF_iij_tau_i}, namely the subtraction of the contributions of the DPLL and pre-DLL low-pass filter. Sweeping the heterodyne beatnote frequencies (as in \cref{fig:het_freq_drift_t}) modulates their amplitudes due to the non-flat response of the analog signal chain. This, in turn, alters the DPLL transfer function and group delays relevant to the PRN signals. \cref{fig:simulated_dll_error_signal_bitbalanced} shows a numerical simulation of the DPLL-induced change of the DLL locking point. Including the pre-DLL low-pass filter in such a simulation, which causes a constant offset, the total contribution $B^{r,j}_{ij,\mrm{DPLL\rightarrow LPF}}$ can be estimated and subtracted from the PRNR observables.

\begin{figure}
    \centering
    \includegraphics[width=8.6cm]{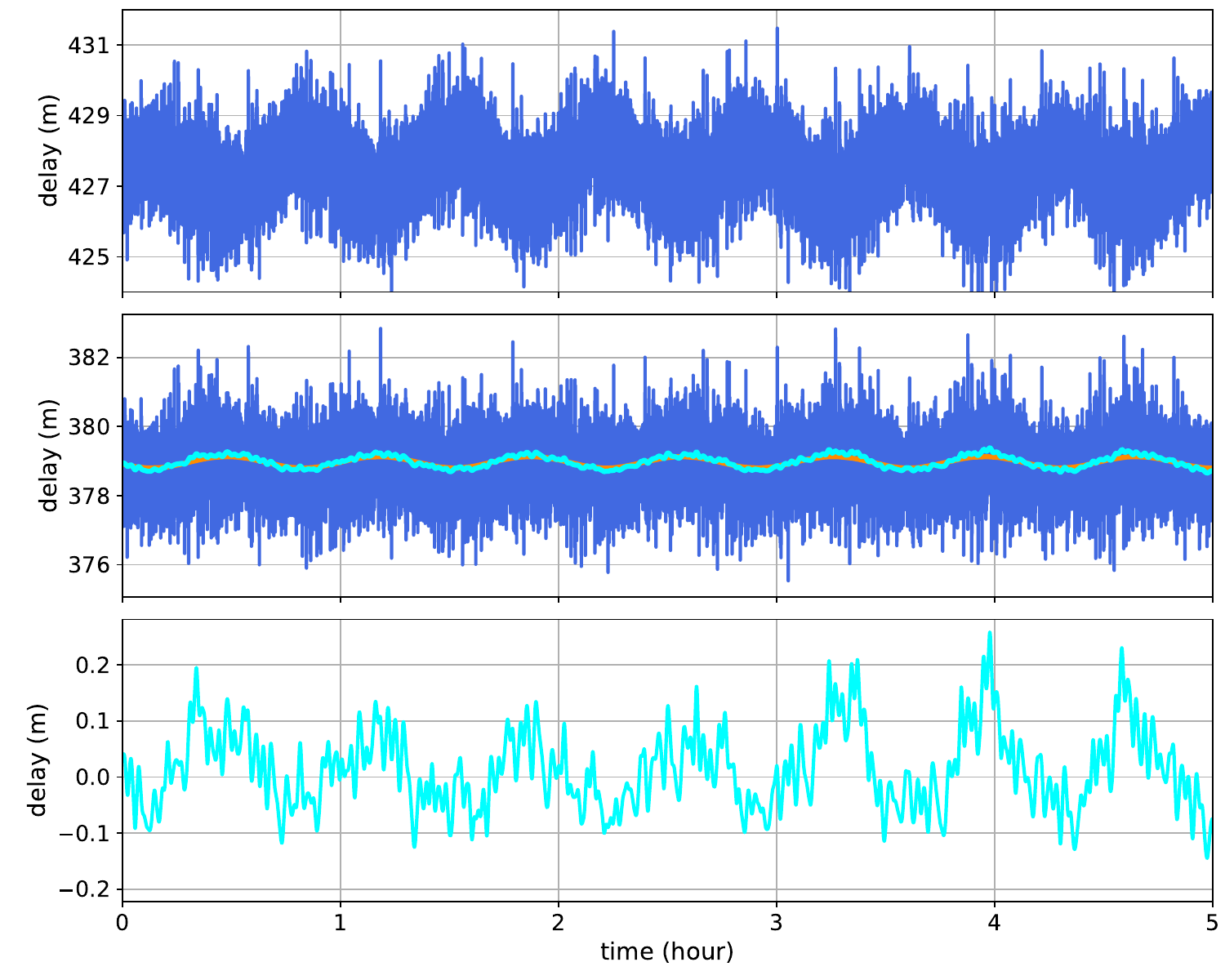}
    \caption{\label{fig:dpll_correction_t} Local PRNR measurement at PM\,1 with a link to PM\,2 $R^{\mrm{prn},\tau_1}_{11,2}$. Top: the PM output without any treatment, namely $R^{\mrm{prn},\tau_1}_{11,2}$ from~\cref{eq:mpr_iij_tau_i}. Middle: blue is the measurement after subtracting the contribution of the DPLL and pre-DLL low-pass filters $R^{\mrm{prn},\tau_1}_{11,2,\mrm{\rightarrow DPLL}}$ from~\cref{eq:Rprn_DPLL_LPF_iij_tau_i}, and cyan is its filtered version. Orange is the delay modulation due to non-constant group delay of the signal chain (e.g., the PR and analog electronics on PM\,1), which was separately calibrated. The signal-chain delay had the mean \SI{1.91}{\meter}, but the offset is adjusted to cyan in this plot. Bottom: the difference between cyan and orange in the middle.}
\end{figure}

\cref{fig:dpll_correction_t} shows this treatment, taking the local PRNR measurement at PM\,1 with a link to PM\,2, $R^{\mrm{prn},\tau_1}_{11,2}$, as an example. This subtraction, from the top to the middle panel, removes the delay modulation caused by the DPLL and the large offset of approximately \SI{50}{\meter}. The residual modulation around \SI{40}{\centi\meter}, recognized in the data filtered by a \SI{10}{\milli\Hz} low-pass filter (cyan), is caused by the transfer function of the signal chain, such as in the PR and analog electronics on PM\,1. The clock synchronization scheme in~\cref{sub:prnr_scheme2} does not correct such a bias; nevertheless, using the separately calibrated transfer function of the signal chain, its contribution to the beatnote bias modulation is projected in orange as a consistency check. This shows the good agreement with cyan, which confirms the treatment via simulation in~\cref{eq:Rprn_DPLL_LPF_ij_tau_i,eq:Rprn_DPLL_LPF_iij_tau_i}. By taking the difference between cyan and orange, $\delta B^{r,i}_{ij,\mrm{BS\rightarrow DPLL}}$ in~\cref{eq:delta_Br_c_ij} is computed in the bottom panel. This mismatch between the beatnote and PRN signals showed the roughly \SI{10}{\centi\meter} residual modulation peak in this particular setup. This is one potential limitation of Scheme 2.

\begin{figure*}
    \centering
    \includegraphics[width=17.2cm]{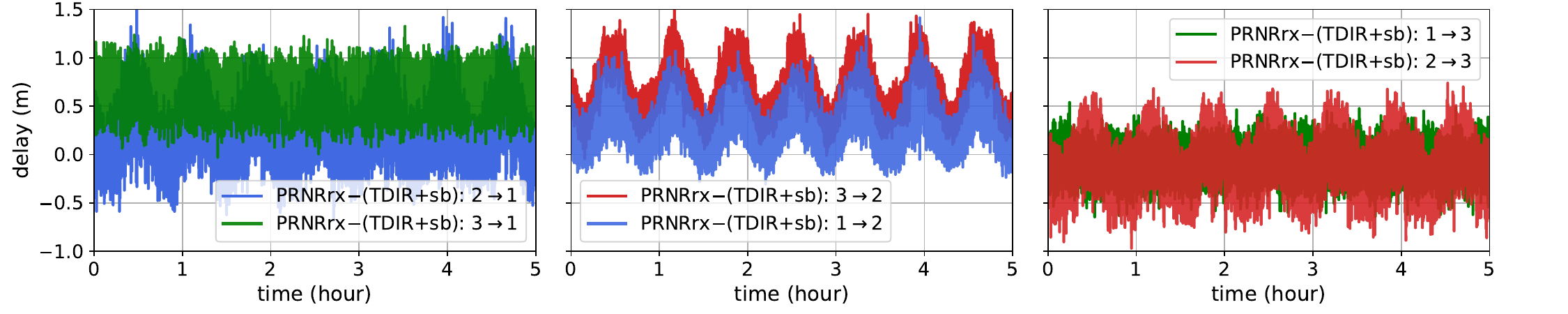}
    \caption{\label{fig:prnr_entire_process_t_2} Differences between the received-PRNR estimates and the combination of TDIR and sidebands in Scheme 2: PM\,1 (left), PM\,2 (middle), PM\,3 (right). Therefore, these are the same quantities as the received PRNR measurements in the third row in \cref{fig:prnr_entire_process_t_1}, but with Scheme 2.}
\end{figure*}

After the simulation-based initial treatment, we performed the three-signal combination with PRNR in Scheme 2 in~\cref{eq:Delta_3PM_tau_1_prnr} for clock synchronization and cyclically repeated it for all three PMs as a reference. \cref{fig:prnr_entire_process_t_2} only shows the differences between received-PRNR and TDIR (+ sideband measurements), same as the third row in~\cref{fig:prnr_entire_process_t_1} in Scheme 1. While TDIR generates the estimates on initial offsets as single values, PRNR continuously tracks the change of delays due to the non-flat group delays. Therefore, the differences still include the time-dependent components and do not directly correspond to PRNR accuracy. That being said, it is remarkable that the values are already in the right range of the classical \SI{1}{\meter} mark. This means the PRNR scheme in~\cref{sub:prnr_scheme2} can achieve this target without the aid of other ranging monitors, other than for ambiguity resolution.

Finally, \cref{tab:resi_tones_2} summarizes the ranging errors derived by comparing the residual tones with the injected ones, as done for Scheme 1. The accuracy with only PRNR is all below \SI{50}{\centi\meter}, which is already below \SI{1}{\meter}, with margin.
The limiting factor is expected to be a relative offset between the local and received PRNR estimates caused by the combination of code interference and ICC~\cite{YamamotoPhD}.
As mentioned in~\cref{sub:testbed_capability}, ICC is feasible only for DLL$_\mrm{RX}$.
The study in~\cite{YamamotoPhD} showed that this readout asymmetry between DLL$_\mrm{RX}$ and DLL$_\mrm{LO}$ resulted in a relative offset between them, around \SI{67}{\centi\meter} over a bidirectional link (i.e., two links between two PMs) in~\cref{fig:bidirectional_setup}. This offset must be studied further.
Some of the other expected main contributors would be the accuracy of the simulation-based initial treatment and the difference in the non-constant group delays between the beatnote and PRNR signals, namely $\delta B^{r,i}_{ij,\mrm{BS\rightarrow DPLL}}$ in~\cref{eq:delta_Br_c_ij}, that highly depends on the PRs' specifications. The latter showed, for example, the potential worst-case offset of \SI{20}{\centi\meter} as shown in the bottom panel in~\cref{fig:dpll_correction_t}.
The accuracy with TDIR calibration (right-most column of~\cref{tab:resi_tones_2}) demonstrates improvements, although it does not yet match the \SI{2.31}{\milli\meter} accuracy achieved with tone-assisted TDIR.
The link between PMs 1 and 3, which had the smallest heterodyne frequency sweep, showed the best performance of around \SI{2.5}{\centi\meter} with TDIR calibration.
This suggests the limited improvement is attributable to the calibration mode performed with the heterodyne frequencies around the edge of the frequency modulation ranges in science mode, as shown in~\cref{fig:het_freq_drift_t}.
For example, $\delta B^{r,i}_{ij,\mrm{BS\rightarrow DPLL}}$ is a function of beatnote frequencies as shown in~\cref{fig:dpll_correction_t}; hence, its coupling to the performance with TDIR calibration was maximized due to the choice of frequencies.
This point gives insight into the lifetime of a single TDIR calibration. Although heterodyne frequencies sweep much slower in LISA, the TDIR calibration values are expected to become inaccurate after some time due to the similar effect. In summary, this issue could be a downside of Scheme 2 compared with Scheme 1, where the biases have all been corrected, and the final performance with TDIR was, on average, better.

\begin{table}[b]
    \caption{\label{tab:resi_tones_2}
    The results of absolute ranging in the three-signal combination at each PM in Scheme 2. Same as Scheme 1 in~\cref{tab:resi_tones_1}, the accuracy was computed from the injected and residual tones.}
    \begin{ruledtabular}
    \begin{tabular}{ccc}
    Parameter & Only PRNR (\si{\centi\meter}) & With TDIR (\si{\centi\meter})\\
    \hline \\\vspace{1.5mm}
    $\text{PM}2\rightarrow \text{PM}1$: $\D^{\mrm{prn},\tau_1}_{12}$ & 18.2 & 13.4\\\vspace{1.5mm}
    $\text{PM}3\rightarrow \text{PM}1$: $\D^{\mrm{prn},\tau_1}_{13}$ & 41.1 & 2.74\\\vspace{1.5mm}
    $\text{PM}3\rightarrow \text{PM}2$: $\D^{\mrm{prn},\tau_2}_{23}$ & 35.5 & 18.5\\\vspace{1.5mm}
    $\text{PM}1\rightarrow \text{PM}2$: $\D^{\mrm{prn},\tau_2}_{21}$ & 24.6 & 18.5\\\vspace{1.5mm}
    $\text{PM}1\rightarrow \text{PM}3$: $\D^{\mrm{prn},\tau_3}_{31}$ & 5.78 & 2.55\\\vspace{1.5mm}
    $\text{PM}2\rightarrow \text{PM}3$: $\D^{\mrm{prn},\tau_3}_{32}$ & 5.86 & 11.3
    \end{tabular}
    \end{ruledtabular}
\end{table}

\section{Conclusion}\label{sec:conclusion}
This article reports the experimental end-to-end demonstration of absolute ranging for LISA using the hexagonal optical bench at the AEI Hannover~\cite{Schwarze2019}. The setup integrates all auxiliary functions~\cite{Heinzel2011} and can generate sets of LISA-like signals featuring six laser links without long-arm distances to simulate intersatellite light travel time.
This study established the experimental base of absolute ranging with PRN modulations for LISA, containing all PRNR treatments, some of which have been discussed analytically and numerically for the actual LISA use case~\cite{Reinhardt2023,Euringer2023B}.

We presented the two PRNR processing schemes (\cref{sub:demo_prnr_1,sub:demo_prnr_2}), requiring various PRNR treatments: unwrapping, ambiguity resolution, bias subtraction, jitter reduction, and signal chain calibration.
As these technologies and PRNR signal processing are applicable to LISA, a similar post-processing scheme is expected to be feasible for the mission.

Scheme 1 aligns with the classical view of PRNR: it measures additional delays, namely biases, in the pseudoranges~\cite{Reinhardt2023}. Therefore, the beatnote and PRNR biases have been corrected for this purpose. The former in~\cref{eq:Acij_nucij} lets the pseudoranges become proper inputs for clock synchronization, and the latter extracts the pseudoranges from PRNR. After all the treatments, as summarized in~\cref{tab:resi_tones_1}, the PRNR performance in Scheme 1 reached around \SI{1}{\centi\meter} to \SI{8}{\centi\meter}, well below the \SI{1}{\meter} requirement~\cite{LISARedBook}, with beatnote frequency drifts on the order of \si{\mega\Hz} over the LISA heterodyne bandwidth. The residual PRNR-bias modulations and the modeling error of the transfer functions from the PRs to the DPLLs limit the performance.

Departing from the pseudoranges, Scheme 2 was also developed based on the latest understanding of the actual needs of TDI~\cite{Reinhardt2023}. The significance of this scheme is that it does not rely on any reference measurement, except for ambiguity resolution, to meet the \SI{1}{\meter} LISA requirement. Therefore, it would be a decent way of using PRNR in a stand-alone use case. As summarized in~\cref{tab:resi_tones_2}, the performance reached \SI{5}{\centi\meter} to \SI{41}{\centi\meter}. It is expected to be limited by the relative offset between the local and received PRNR estimates due to the combination of code interference and ICC~\cite{YamamotoPhD}. Scheme 2 did not show the expected improvement with TDIR calibration. This is expected to be caused by the selected heterodyne frequencies in calibration mode from the edge frequency regions of science mode.
In this sense, Scheme 2 might require better care in the operation of calibration mode compared with Scheme 1 because of its potential limiting factor: the difference in the non-constant group delays between the beatnote and PRNR signals $\delta B^{r,i}_{ij,\mrm{BS\rightarrow DPLL}}$ in \cref{fig:dpll_correction_t}. Because of its dependency on the heterodyne frequency, we should use the calibration value associated with the average heterodyne frequency to analyze. With such treatment, Scheme 2 is expected to overcome the residual biases more than \SI{10}{\centi\meter}, shown in the right column in \cref{tab:resi_tones_2}.
Also, the application of Scheme 2 to the LISA interferometric topology with onboard delays~\cite{Euringer2023B,Reinhardt2024}, where we have more local interferometers, is being studied by some of the authors.

These PRNR processing schemes rely on local PRNR for both bias estimation and stochastic jitter suppression via ICC, as in \cite{Sutton2013}. ICC with data encoding requires an additional internal feed-forward within the local PM. While ICC also impacts data communication performance, this aspect is beyond the scope of this article.

In the authors' view, the capability of tone injections via laser-lock loops (cavity lock and/or offset lock) is very useful in LISA. As demonstrated in this article, the tones can enhance and/or calibrate ranging accuracy. They can be activated at least during the in-flight debugging mode.

Further tasks to be considered will be listed here.
First, the weak-light condition is expected to affect in-band ranging jitters but not biases. Nevertheless, it is essential to implement a weak light experiment as a technology demonstrator for the LISA mission.
The Hexagon is being developed in this direction.
Second, this study did not investigate thermally induced drifts of biases. For example, the EOM would be expected to be the biggest contributor to the transmitter bias drift. Scheme 2 cancels the transmitter bias by its algorithm, while Scheme 1 is affected by such a drift.
Third, as highlighted in~\cref{fig:simulated_dll_error_signal_bitbalanced,fig:dpll_correction_t}, the non-flat-magnitude responses of components in the signal chain convert the drift of the heterodyne frequency into the signal amplitude modulation, which results in the modulation of the DPLL group delay~\cite{Gerberding2013}.
This coupling is expected to be significantly suppressed by an automatic gain control system, which could ease PRNR post-processing.
We are also considering implementing such a gain control system or similar.
Fourth, it would be possible to resolve the ambiguity of sideband phases using PRNR, which is the order of \SI{10}{\centi\meter}~\cite{Reinhardt2023}.
In this case, the ``ranging ladder" could be deepened by one more level: ground observation (or TDIR), PRNR, and sideband phases. Correcting biases of sideband phases against PRNR would be highly challenging.
Fifth, identifying individual contributors in the setup to the local PRNR measurements would advance confidence in absolute ranging.
Sixth and last, it might be possible to mimic some aspects of the impact from LISA-like intersatellite light travel time. Lacking the long-arm entity in the Hexagon is not a significant shortcoming as a technology demonstrator because we can test the complete PM transmitter and receiver (see \cref{fig:pm_entire_architecture}), where most of the PRNR biases occur, and the clocks contribute. Also, the weak light mentioned above would cover the most important aspect of the SC distances concerning PRNR. Nevertheless, some features are still missing and would be interesting to test, e.g., the disentanglement of the onboard clocks and the intersatellite delays~\cite{Reinhardt2024b}. To demonstrate it, for example, we could coherently inject an electronic signal that simulates the time-variant SC separation to the clocks, the PRN signals, and the carrier via laser locks.

\begin{acknowledgments}
The authors thank Martin Hewitson and Diogo Coutinho for useful discussions. The authors acknowledge support by the German Aerospace Center (DLR) with funds from the Federal Ministry of Economics and Technology (BMWi) according to a decision of the German Federal Parliament (Grant No. 50OQ2301, based on Grants No. 50OQ0601, No. 50OQ1301, No. 50OQ1801).
The authors also acknowledge support by the Deutsche Forschungsgemeinschaft (DFG, German Research Foundation) Project-ID 434617780-SFB 1464, and the Deutsche Forschungsgemeinschaft (DFG) Sonderforschungsbereich 1128 Relativistic Geodesy and Cluster of Excellence “QuantumFrontiers: Light and Matter at the Quantum Frontier: Foundations and Applications in Metrology” (EXC-2123, Project No. 390837967). Olaf Hartwig and Martin Staab gratefully acknowledge support from the Centre National d’\'Etudes Spatiales (CNES). Jan Niklas Reinhardt acknowledges the funding by the Deutsche Forschungsgemeinschaft (DFG, German Research Foundation) under Germany's Excellence Strategy within the Cluster of Excellence PhoenixD (EXC 2122, Project ID 390833453).

\end{acknowledgments}

\appendix

\section{Tone-assisted TDIR}\label{sec:tone_tdir}

Secondary noises limit the accuracy of TDIR, as was the case in \cite{Yamamoto2022}. Tone-assisted TDIR~\cite{Mitryk2012,Francis2015} improves the ranging accuracy by intentionally injecting a sinusoidal tone in the high-frequency region of the PM measurement band,
\begin{align}
	\nu_{T}(\tau) &= |\nu_{T}|\sin(2\pi f_{T}\tau),
	\label{eq:tone}
\end{align}
for example at $f_{T}=\SI{1}{\Hz}$. The tone can be applied to an offset frequency in a laser-lock loop as depicted in \cref{fig:experimental_setup}. This is then distributed via laser locks throughout the system. This enhances the signal-to-noise ratio (SNR) of TDIR, and only noises at the particular tone frequencies are analyzed in post-processing. One critical constraint is that the tones must be injected to enhance all target physical quantities of TDIR via the tones in a final signal combination without a degeneracy with another target quantity.

Tones can be injected to reference frequencies of offset frequency locks for secondary lasers (i.e., 2 and/or 3) in the Hexagon. In this case, three beam frequencies are expressed in an arbitrary time frame as follows,
\begin{align}
	\nu_{c,1}(\tau) &= \nu_{0} + \delta\nu_{1}(\tau),
	\label{eq:beam_1_tdir}\\
	\nu_{c,2}(\tau) &= \nu_{c,1}(\tau) + O_{12}(\tau) + \delta\nu_{12}(\tau) + \nu_{T2}(\tau),
	\label{eq:beam_2_tdir}\\
	\nu_{c,3}(\tau) &= \nu_{c,1}(\tau) + O_{13}(\tau) + \delta\nu_{13}(\tau) + \nu_{T3}(\tau),
	\label{eq:beam_3_tdir}
\end{align}
where $O_{ij}$ is a \si{\mega\hertz} offset frequency, $\delta\nu_{ij}(\tau)$ is a residual stochastic frequency noise, and $\nu_{Tj}(\tau)$ is an injected frequency tone. These single laser frequencies are mixed via the Hexagon interferometer and result in three beatnotes,
\begin{align}
	\nu_{c,12}(\tau) &= O_{12}(\tau) + \delta\nu_{12}(\tau) + \nu_{T2}(\tau), \label{eq:beatnote_12_tdir}
	\\
	\nu_{c,23}(\tau) &= O_{13}(\tau)-O_{12}(\tau) + \delta\nu_{13}(\tau)-\delta\nu_{12}(\tau)
    \nonumber\\
    &\hspace{15mm}+ \nu_{T3}(\tau)-\nu_{T2}(\tau), 
	\label{eq:beatnote_23_tdir}
	\\
	\nu_{c,31}(\tau) &= - O_{13}(\tau) - \delta\nu_{13}(\tau) - \nu_{T3}(\tau).
	\label{eq:beatnote_31_tdir}
\end{align}

Combining two carrier-carrier beatnotes locally at each PM, the three signal combination with the primary PM $i$ can be expressed in an arbitrary form with small errors of timing estimation $\delta\hat\tau_{j,\mrm{err}}$ and $\delta\hat\tau_{k,\mrm{err}}$,
\begin{align}
	\Delta^{\tau_{i}}_\mrm{3PM}(\tau) &= \nu^{\tau_{i}}_{c,jk}(\tau)+\nu^{\tau_{i}}_{c,ki}(\tau+\delta\tau_{j,\mrm{err}})+\nu^{\tau_{i}}_{c,ij}(\tau+\delta\tau_{k,\mrm{err}})
	\nonumber\\
	&\approx \dot{\nu}^{\tau_{i}}_{c,ki}(\tau)\cdot\delta\tau_{j,\mrm{err}} + \dot{\nu}^{\tau_{i}}_{c,ij}(\tau)\cdot\delta\tau_{k,\mrm{err}}.
	\label{eq:3s_ijk}
\end{align}

Hence, the tones lift the timing-error couplings in different manners, depending on reference time frames,
\begin{align}
	\Delta^{\tau_{1}}_\mrm{3PM}(\tau) &\approx \dot{\nu}^{\tau_1}_{T2}(\tau)\cdot\delta\tau^{\tau_1}_{3,\mrm{err}} - \dot{\nu}^{\tau_1}_{T3}(\tau)\cdot\delta\tau^{\tau_1}_{2,\mrm{err}},
	\label{eq:3s_tau_1}\\
	\Delta^{\tau_{2}}_\mrm{3PM}(\tau) &\approx \dot{\nu}^{\tau_2}_{T2}(\tau)\cdot\left(\delta\tau^{\tau_2}_{3,\mrm{err}} - \delta\tau^{\tau_2}_{1,\mrm{err}}\right) + \dot{\nu}^{\tau_2}_{T3}(\tau)\cdot\delta\tau^{\tau_2}_{1,\mrm{err}},
	\label{eq:3s_tau_2}\\
	\Delta^{\tau_{3}}_\mrm{3PM}(\tau) &\approx -\dot{\nu}^{\tau_3}_{T2}(\tau)\cdot\delta\tau^{\tau_3}_{1,\mrm{err}} +  \dot{\nu}^{\tau_3}_{T3}(\tau)\cdot\left(\delta\tau^{\tau_3}_{1,\mrm{err}} - \delta\tau^{\tau_3}_{2,\mrm{err}}\right),
	\label{eq:3s_tau_3}
\end{align}
where the stochastic frequency noise terms $\delta\nu_{ij}$ are omitted, focusing on the tones. \crefrange{eq:3s_tau_1}{eq:3s_tau_3} suggests that the two tones $\nu_{T2}$ and $\nu_{T3}$ must be at different frequencies; otherwise, two timing errors degenerate or one of them cannot be lifted by the tones.

\begin{figure}
    \centering
    \includegraphics[width=8.6cm]{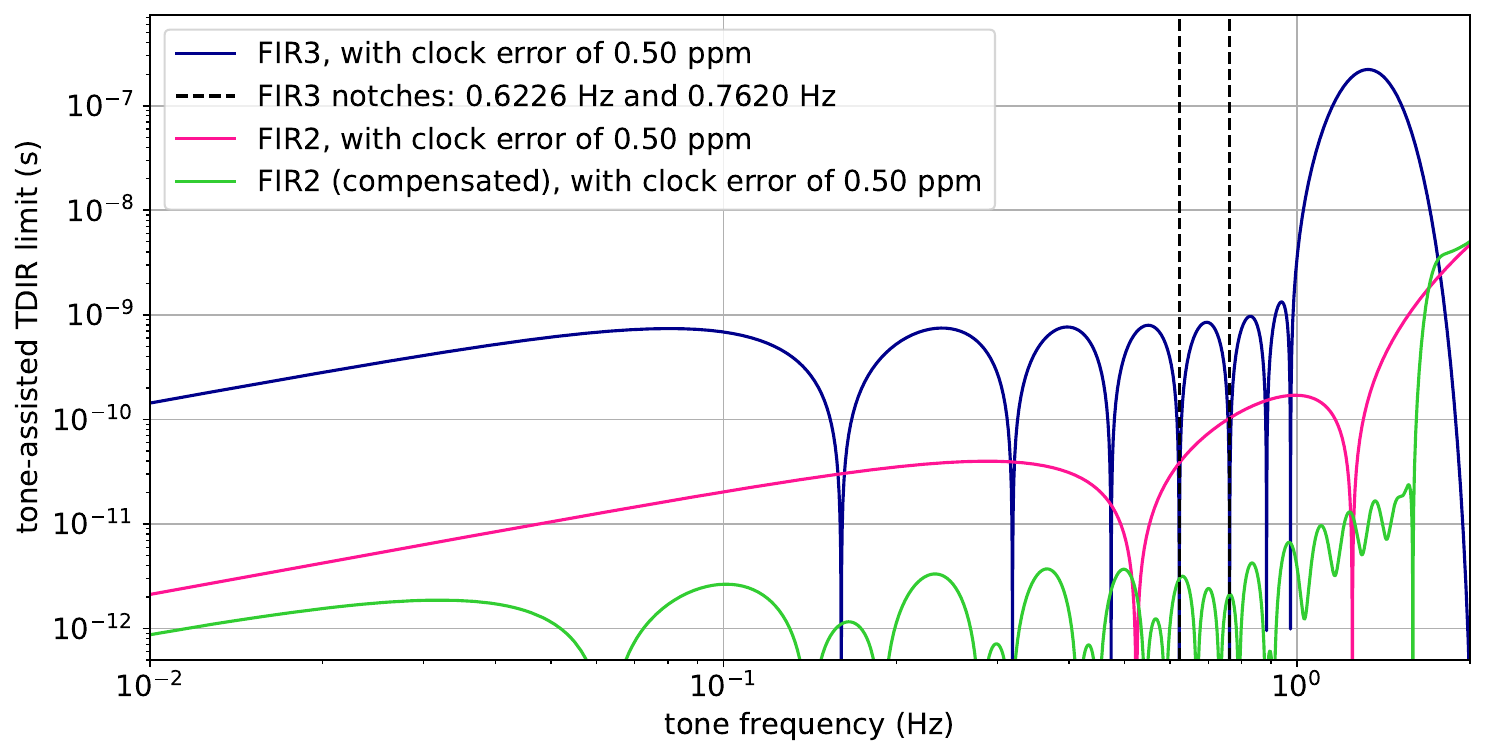}
    \caption{\label{fig:tone_tdir_limit_ff} Flexing-filtering coupling due to each FIR-filter decimation stage: navy, magenta and light-green are the third FIR filter, the second FIR filter, and the compensated second FIR filter, respectively.}
\end{figure}

Tone-assisted TDIR is ideally limited by a stochastic secondary noise floor. However, some technical limitations to laser noise suppression, such as the phase-locking errors and the flexing-filtering coupling~\cite{Bayle2019,Staab2023}, scale with the input beatnote noises; hence, no more accuracy improvement is expected once they become dominant and leave residual tones. After clock synchronization, the former cancels in a three-signal combination in the Hexagon, while the latter does not. Therefore, the tone frequencies were selected to minimize the flexing-filtering coupling. The flexing-filtering coupling from the second and third FIR filters is plotted in \cref{fig:tone_tdir_limit_ff}; see \cite{YamamotoPhD} for its formulation. The idea is to place tones at notches of the transfer function of the third and dominant FIR filter (navy), e.g., \SI{0.6226}{\Hz} and \SI{0.7620}{\Hz}. In this case, the second FIR filter becomes dominant at the tone frequencies (magenta), which can be further suppressed by applying the quasi-inverse filter in post-processing (light-green). The flexing filtering coupling is expected to be suppressed to a few pico-second, or equivalently milli-meter, contribution over the averaging time of \OSI{10000}{\second}.

\begin{figure}
    \centering
    \includegraphics[width=8.6cm]{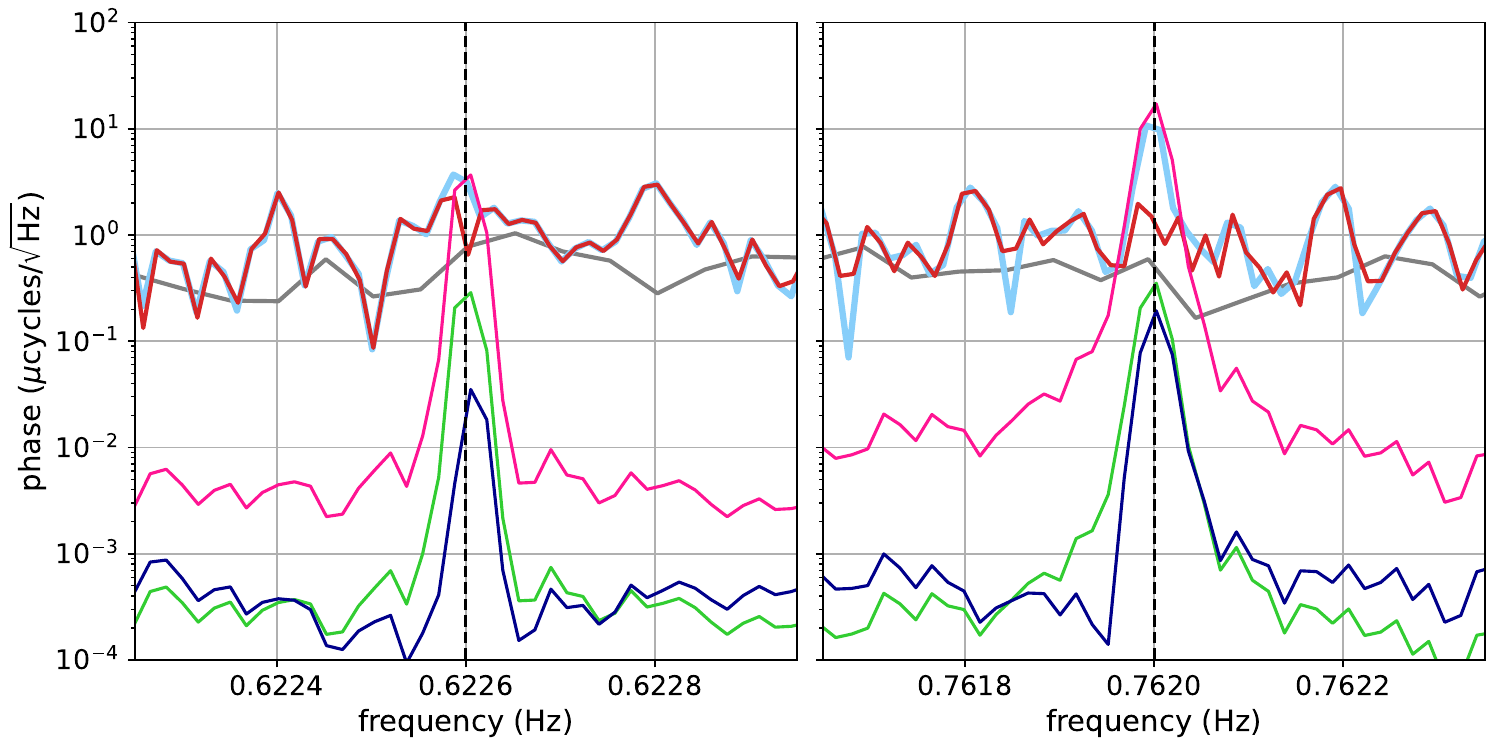}
    \caption{\label{fig:tone_tdir_zoom_f_sci} Clock synchronization $\Delta^{\tau_{1}}_\mrm{3PM}$ in science mode via tone-assisted TDIR, with and without post-processing compensation for the second FIR filter. Red and grey: identical to the ones in \cref{fig:prnr_sync_f_1}. Light-blue: three-signal combination without the post-processing compensation for the flexing-filtering coupling from the second FIR filter. Navy, magenta and light-green: flexing-filtering couplings from different FIR filters, i.e., the third FIR filter, the second FIR filter, and the compensated second FIR filter, respectively.}
\end{figure}

\cref{fig:tone_tdir_zoom_f_sci} shows the same measurement in science mode as \cref{fig:prnr_sync_f_1} but focuses on tone-assisted TDIR and the flexing-filtering couplings. 
A three-signal combination without the post-processing compensation in light-blue reveals the residual tones due to the second FIR filter in magenta as expected. As shown in red, which is identical to the one in \cref{fig:prnr_sync_f_1}, such a compensation suppressed the tones down to a secondary noise floor. In this case, the accuracy of the tone TDIR in this experiment can be written with an averaging time $T_\mrm{avg}$,
\begin{align}
	\delta\tau_{\mrm{err}} = \frac{\SI{2.037e-9}{}}{\sqrt{T_\mrm{avg}}}\hspace{1mm}\text{(s)},
	\label{eq:tone_TDIR_accuracy}
\end{align}
where the testbed noise floor around tone frequencies of \SI{0.550}{\micro{cycles}\prtHz}, a data rate of \SI{3.391}{Sps}, and a tone amplitude of \SI{1}{\kilo\Hz} are absorbed into the value in the equation.

\section{Relation to TDI and symmetries of delay operators}\label{sec:tdi}
To more closely link the Hexagon setup to LISA, we can formally re-write the measurement equations from~\cref{eq:nu_tau_m_ij} in the form
\begin{equation}
    \nu_{c,ij}^{\tau_i} = \dot{D}_{ij}\nu_{c,j}^{\tau_j} - \nu_{c,i}^{\tau_i},
\end{equation}
where we express each laser noise term in the time frame of the associated PM clock. I.e., $\dot{D}_{ij}$ is the operator to synchronize frequency measurements from PM $j$ to PM $i$. This gives us 6 equations with the same algebraic structure as the corresponding variables usually used as a starting point for TDI in the full LISA setup~\cite{Tinto:2020fcc}. 

Consequently, any data combination that cancels laser noise for the full LISA setup (with 6 unique delays) should also work in the Hexagon.

However, while this way of writing the data combinations allows a direct mapping to LISA TDI, it is rather artificial. The delays in the Hexagon have some extra symmetries, which imply that certain (much simpler) data combinations are laser-noise-free in the Hexagon but not in the real LISA setup. Since we don't have any physical delays between the PMs, any sequence of delays that starts and ends at the same PM will be trivial. I.e., we have $\dot{D}_{i_1i_2...i_k} = 1$ for any combinations of delays with $i_1 = i_k$.

It is, in fact, easy to see that the equivalent to the TDI space for the Hexagon experiment has just dimension 1. Formulating the problem in a global frame, as in \cref{eq:nu_tau_m_ij}, and using the symmetry $\nu_{c,ij}^{\tau_m} = -\nu_{c,ji}^{\tau_m}$ reduces the problem to a set of 3 equations. Following~\cite{Vallisneri:2020otf}, the laser-noise-free TDI space can be identified as the null space of the matrix mapping the three laser noise terms into the measurements, which in our case can be formulated at a single instance of time. I.e., we're looking for a matrix $T$ such that
\begin{equation}
\begin{split}
   T \cdot &\begin{pmatrix}
 \nu_{c,12}^{\tau_m}\\
 \nu_{c,23}^{\tau_m} \\
 \nu_{c,31}^{\tau_m}
\end{pmatrix} = T \cdot \underbrace{\begin{pmatrix}
 -1 & 1 & 0 \\
 0 & -1 & 1 \\
 1 & 0 & -1 \\
\end{pmatrix}}_{M} \cdot \begin{pmatrix}
 \nu_{c,1}^{\tau_m}\\
 \nu_{c,2}^{\tau_m} \\
 \nu_{c,3}^{\tau_m}
\end{pmatrix} = 0\quad \forall \nu_{c,1}^{\tau_m}, \nu_{c,2}^{\tau_m},\nu_{c,3}^{\tau_m},
\\
&\implies T M = 0.
\end{split}
\end{equation}
This implies the rows of $T$ are giving a basis of the null space of $M^T$, which has the unique solution given by \cref{eq:Delta_1PM_tau_m}. This implies any 6 link TDI solution should, after synchronizing the measurements and eliminating combinations of the form $\nu_{c,ij}^{\tau_m} + \nu_{c,ji}^{\tau_m}$, reduce to linear combinations of the fundamental 3 signal combination.

\section{LISA-like signal combinations}\label{sec:lisa_combi}
The signal combination applied in this article in \cref{eq:Delta_3PM_tau_i} is straightforward but not best in line with LISA signal processing. This section provides information on LISA-like signal combinations that are configurable with the Hexagon for testing the LISA data analysis pipeline with its real data.

Combinations in analogy to the Michelson TDI combinations  $X$, $Y$, and $Z$ are, in principle, just the reduced version of \cref{eq:Delta_3PM_tau_i},
\begin{align}
    X^{\tau_i}(\tau) &= \nu^{\tau_i}_{c,ij}(\tau) + \dot{D}^{\tau_i}_{ij}\nu^{\tau_j}_{c,ji}(\tau) - \nu^{\tau_i}_{c,ik}(\tau) - \dot{D}^{\tau_i}_{ik}\nu^{\tau_k}_{c,ki}(\tau),
    \label{eq:tdi_xyz}
\end{align}
where $\dot{D}_{ij}$ represents a delay operator which transforms time frames from clock $j$ to clock $i$ for a beatnote frequency. This combination, however, is against the original motivation of the Hexagon as the PM testbed, namely the optical three-signal test to probe the PM nonlinearity~\cite{Schwarze2019}. Indeed, applying the time delays in \cref{eq:tdi_xyz} expresses the laser noise terms in a common frame, in which the first two summands and the last two summands already cancel independently. 

To overcome the issue with TDI $X$, $Y$, and $Z$ above, the Sagnac combination would be the possible best LISA-like combination performed in the Hexagon,
\begin{align}
    \alpha^{\tau_1}(\tau) &= \nu^{\tau_1}_{c,12}(\tau) + \dot{D}^{\tau_1}_{12}\nu^{\tau_2}_{c,23}(\tau) + \dot{D}^{\tau_1}_{12}\dot{D}^{\tau_2}_{23}\nu^{\tau_3}_{c,31}(\tau)
    \nonumber\\
    &\hspace{3mm}-\left(\nu^{\tau_1}_{c,13}(\tau) + \dot{D}^{\tau_1}_{13}\nu^{\tau_3}_{c,32}(\tau) + \dot{D}^{\tau_1}_{13}\dot{D}^{\tau_2}_{32}\nu^{\tau_2}_{c,21}(\tau)\right).
    \label{eq:sagnac_alpha}
\end{align}
Each of the first and second lines in \cref{eq:sagnac_alpha} configures a three-signal combination and therefore is noise-free; however, importantly, combining the two three-signal combinations benefits between balanced detection from a pair of $\nu_{ij}$ and $\nu_{ji}$, same as \cref{eq:Delta_3PM_tau_i}. This is quite similar to the three-signal combination $\Delta^{\tau_{i}}_\mrm{3PM}$ in~\cref{eq:Delta_3PM_tau_i}. But they are different in that $\alpha^{\tau_1}$ requires four pseudoranges instead of two.

Finally, we want to mention the $\zeta$ combination, which together with Sagnac combination $\alpha$ and its two cyclic permutations $\beta,\gamma$ forms a generating set of the TDI space for LISA under the assumption of 3 constant but unequal arms~\cite{Dhurandhar:2001tct}. However, this combination is usually only defined in the setup of a static LISA constellation with synchronized clocks, in which $\dot{D}^{\tau_i}_{ij} = \dot{D}^{\tau_j}_{ji}$ holds and there are only 3 unique delays. In that case, it can be written as a fully symmetric expression
\begin{equation}
\begin{split}
    \zeta = &\dot{D}^{\tau_2}_{23}(\nu^{\tau_1}_{c,12}(\tau) - \nu^{\tau_1}_{c,13}(\tau))
    \\
    + &\dot{D}^{\tau_3}_{31}(\nu^{\tau_2}_{c,23}(\tau) - \nu^{\tau_2}_{c,21}(\tau)) 
    \\
    + &\dot{D}^{\tau_1}_{12}(\nu^{\tau_3}_{c,31}(\tau) - \nu^{\tau_3}_{c,32}(\tau)).
\end{split}
\end{equation}
In our case, we instead have $\dot{D}^{\tau_i}_{ij} = (\dot{D}^{\tau_j}_{ji})^{-1}$, such that this version of $\zeta$ does not properly synchronize the clocks and does not cancel the laser noise. Had we instead first synchronized the PM measurements independently, we could remove all delays $\dot{D}^{\tau_i}_{ij}$ in the above expression, in which case $\zeta$ again reproduces the three-signal combination with balanced detection.

\bibliography{reference}

\end{document}